\documentclass[conference]{IEEEtran}
\IEEEoverridecommandlockouts
\usepackage{cite}
\usepackage{amsmath,amssymb,amsfonts}
\usepackage{mathrsfs} 
\usepackage{algorithmic}
\usepackage{graphicx}
\usepackage{booktabs} 
\usepackage{caption}
\usepackage{enumitem}
\setenumerate[1]{itemsep=0pt,partopsep=0pt,parsep=\parskip,topsep=5pt}
\setitemize[1]{itemsep=0pt,partopsep=0pt,parsep=\parskip,topsep=5pt}
\setdescription{itemsep=0pt,partopsep=0pt,parsep=\parskip,topsep=5pt}
\usepackage{bm}
\usepackage{textcomp}
\usepackage{xcolor}
\usepackage{subfigure}                
\def\BibTeX{{\rm B\kern-.05em{\sc i\kern-.025em b}\kern-.08em
    T\kern-.1667em\lower.7ex\hbox{E}\kern-.125emX}}
\begin{document}

\title{Model-Driven Based Deep Unfolding Equalizer for Underwater Acoustic OFDM Communications}
\author{Hao Zhao, Cui Yang, Yalu Xu, Fei Ji,~\IEEEmembership{Member,~IEEE},\\    Miaowen Wen,~\IEEEmembership{Senior Member,~IEEE}, and Yankun Chen
 \thanks{ Hao Zhao, Cui Yang, Yalu Xu, Fei Ji, and  Miaowen Wen are with School of Electronic and Information Engineering, South China University of Technology, Guangzhou, 510640, China, and also with key Laboratory of Marine Environmental Survey Technology and Application, Ministry of Natural Resources, Guangzhou, 510300, China (e-mail:  ctzhaohao@mail.scut.edu.cn, yangcui@scut.edu.cn, eexyl@mail.scut.edu.cn,
, eefeiji@scut.edu.cn,  eemwwen@scut.edu.cn).



Yankun Chen is with Key Laboratory of Marine Environmental Survey Technology and Application, Ministry of Natural Resources, Guangzhou, 510300, China (e-mail:easiechen@qq.com).

 }
}
\maketitle

\begin{abstract}

\textcolor{black}{It is challenging to design an equalizer for the complex time-frequency doubly-selective channel. In this paper, we employ the deep unfolding approach to establish an equalizer for the underwater acoustic (UWA) orthogonal frequency division multiplexing (OFDM) system, namely UDNet. Each layer of UDNet is designed according to the classical minimum mean square error (MMSE) equalizer. Moreover, we consider the QPSK equalization as a four-classification task and adopt minimum Kullback-Leibler (KL) to achieve a smaller symbol error rate (SER) with the one-hot coding instead of the MMSE criterion. In addition, we introduce a sliding structure based on the banded approximation of the channel matrix to reduce the network size and aid UDNet to perform well for different-length signals without changing the network structure.
Furthermore, we apply the measured at-sea doubly-selective UWA channel and offshore background noise to evaluate the proposed equalizer. Experimental results show that the proposed UDNet performs better with low computational complexity. Concretely, the SER of UDNet is nearly an order of magnitude lower than that of MMSE.}


\end{abstract}

\section{Introduction}

\textcolor{black}{Underwater acoustic (UWA) communications is one of the most effective ways to achieve underwater information exchange because of its low attenuation characteristic compared to
	the electromagnetic wave and light wave\cite{acousticAdvantage}. 
	However, UWA communications face more severe difficulties than wireless communications,   i.e., the frequency-dependent attenuation, the time-varying multipath, and the severe Doppler effect\cite{b2,b3}. 
	Moreover, the underwater scenario has rich interference, which also brings obstacles to the UWA communications, including the noise from sonar operations, marine mammals, shrimp snapping noise, and malicious jamming\cite{noise11}\cite{noise12}. It is generally recognized as one of the most challenging communication tasks\cite{difficulty}.}  
Fortunately, orthogonal frequency division multiplexing (OFDM) as a multi-carrier modulation technique has been proven to be a viable option for achieving high data-rate communications. It has a good performance in dealing with the long-delay multipath UWA channels\cite{OFDM2}. Meanwhile, it has shown robustness against underwater inter-symbol interference (ISI)\cite{OFDM3}. 
However, a practical issue is that doubly-selective channels will cause inter-carrier interference (ICI)~\cite{ofdmICI}.   The ICI degrades OFDM performance, especially for the system employing a large number of sub-carriers in a limited bandwidth or   high-order constellation modulations. 
On the one hand, the UWA channel is wideband due to the small ratio of the carrier frequency to the signal bandwidth, which makes the communication system more sensitive to carrier frequency offset caused by Doppler shifts. On the other hand, the non-uniform doppler also poses a significant challenge for the equalizer design. The non-uniform ICI reduction renders existing equalizers ineffective\cite{NonuniformICI}.
The classical interference cancellation methods, such as zero-forcing (ZF) and minimum mean squared error (MMSE) equalizers, are usually based on the assumption that all sub-carriers experience the same ICI.
The ICI between each sub-carrier is non-uniform, which makes the frequency-domain channel matrix irregular banded and very hard to model. Meanwhile, it brings enormous difficulty to channel equalization and symbol detection. Significantly when channel estimation is imperfect, the performance of the classical equalizer degrades sharply.
\textcolor{black}{
	Besides, plenty of pre-coded  OFDM systems are gradually emerging to improve the quality of underwater information transmission, e.g., vector-OFDM\cite{vectorOFDM}, orthogonal signal division multiplexing (OSDM)\cite{QuOSDM,tspOSDM,hanOSDM}, and orthogonal time frequency space (OTFS)\cite{OTFS}. In literature, the MMSE equalizer performs as an extremely fundamental and essential function in the corresponding equalization procedure, which strikes a good tradeoff between the bit error rate (BER) performance and the computational complexity.}

In recent years, deep learning (DL)  methods have achieved several excellent results in communications. In particular, DL has different applications, such as signal estimation, modulation recognition, and equalization. Li \emph{et al.} first attempted to utilize an end-to-end OFDM receiver as a fully connected deep neural network (FC-DNN) that replaces channel estimation, signal detection, and demodulation modules jointly based on the DNN network~\cite{b14}. The drawback of FC-DNN is that DL-receiver treats itself as a black box, which makes the data-driven method unexplainable and unpredictable. 
To solve these issues, a model-driven DL-receiver called ComNet was proposed, which combines expert knowledge of communication algorithms and exhibits higher data recovery accuracy than FC-DNN\cite{b15}. In the field of UWA signal processing, a DL-based receiver was designed for single-carrier communications over time-varying UWA channels. It works with online training and test modes for accommodating the time variability of UWA channels\cite{b16}. Aiming to improve the OFDM communication performance over UWA doubly-selective channels, a convolutional neural network (CNN) based architecture was proposed to compose an encoder and a decode\cite{b17}.
\textcolor{black}{The drawback of the integrated design is that the DL receiver is sensitive to the data. It has a strong requirement for data matching. The training dataset and the test dataset should be similar or the same.}
\textcolor{black}{Therefore, to maintain the stability of the communication system, some researchers replace the specific signal processing module with a DL network in the communication system.
	For example, the channel estimator was replaced by a super-resolution network to improve the accuracy of channel estimation\cite{SuperEst}. Moreover, integration DNNs into the Viterbi algorithm was proposed to improve symbol detection performance\cite{ViterbiNet}. 
	Recently, the unfolding algorithm is becoming a promising approach to design a DL-based module according to the existing complex algorithms\cite{b20}.
	Based on the current maximum likelihood (ML) algorithm model and deep unfolding method, a detector called DetNet, was constructed for multi-input-multi-output (MIMO) detection\cite{b21}. In this work, the learning begined with the existing algorithm as an initial starting point, and a soft decision output method was proposed. Further, a  sliding cascaded network,  dubbed SCN, was designed for large Doppler frequency shift channels\cite{b13}. The deep unfolding method generally achieves better signal equalization performance than the existing traditional methods, i.e., DetNet and SCN. Unfortunately, both of them still follow the regression idea and use MMSE as the criterion for optimizing network parameters. 
	Moreover, constellation classification has already been studied based on DL\cite{constellation1,constellation2}.
	Motivated by this, a novel unfolding-based equalizer using the conception of classification is proposed in this paper. Compared with existing works, our main contributions are summarized as follows:}

\begin{itemize}
\item \textcolor{black}{We propose a DL equalizer based on an unfolding algorithm called UDNet, to improve the frequency domain equalization performance in complex UWA-OFDM systems. The classification idea is considered for the equalization task. Specifically, the $\kappa$-constellation recognition is regarded as a $\kappa$-category issue instead of a regression problem.	  
	To evaluate the gap between the label and predictions,  we employ the minimum KL criterion. 	The residual connection is employed to accelerate the training convergence. Moreover, we apply a sliding structure that assists UDNet in adapting some length signals without modifying the network architecture. The sliding structure enables UDNet to handle received signals of different sub-carriers, which is more practical.
}

\item \textcolor{black}{To make experiments more convincing,  we adapted the measured at-sea UWA channel impulse response and the additive offshore noise measured in Nanpeng island, Yangjiang, China, to evaluate the performance of the proposed  UDNet.  Experiment  show that UDNet obtains better accuracy than the classical ones, i.e., the semidefinite relaxation equalizer, the decision feedback equalizer, the DetNet detector, and the SCN detector. Meanwhile, the UDNet consumes the minimum computing time under the same computing equipment.}

\end{itemize}

The structure of the rest of  this paper is as follows. Section II establishes the doubly-selective UWA channel models, and both perfect channel state information (CSI) and imperfect CSI are considered. In Section III, the proposed equalizer network structure is introduced in detail. In Section IV, we provide the experimental parameters of the network training and the performance of the proposed network in different channel models. Moreover, our network is tested with the real offshore background noise. The final section concludes the paper.

$\textbf{Notation:}$ In this paper, we define the normal distribution whose mean value is ${u}$ and standard deviation is $\sigma$ as $\mathcal N({u},\sigma)$. We utilize  uppercase letters to represent variables in the frequency domain and we use lowercase letters to represent  variables in the time domain. Meanwhile,  we employ boldface  letters to represent matrices and vectors. The superscript $(\cdot)^{T}$ means the transpose. The $i$-th element of the vector ${s}$ is denoted as ${s}_{i}$. The $\mathfrak{R}(\cdot)$ represents the real part, and the $\mathfrak{I}(\cdot)$ represents the imaginary part when considering a complex vector or matrix. 

\section{SYSTEM MODEL}
\subsection{UWA-OFDM System}\label{AA}
\begin{figure}[tb!]
	\captionsetup{font={small}}
	\centering 
	\includegraphics[width=10cm]{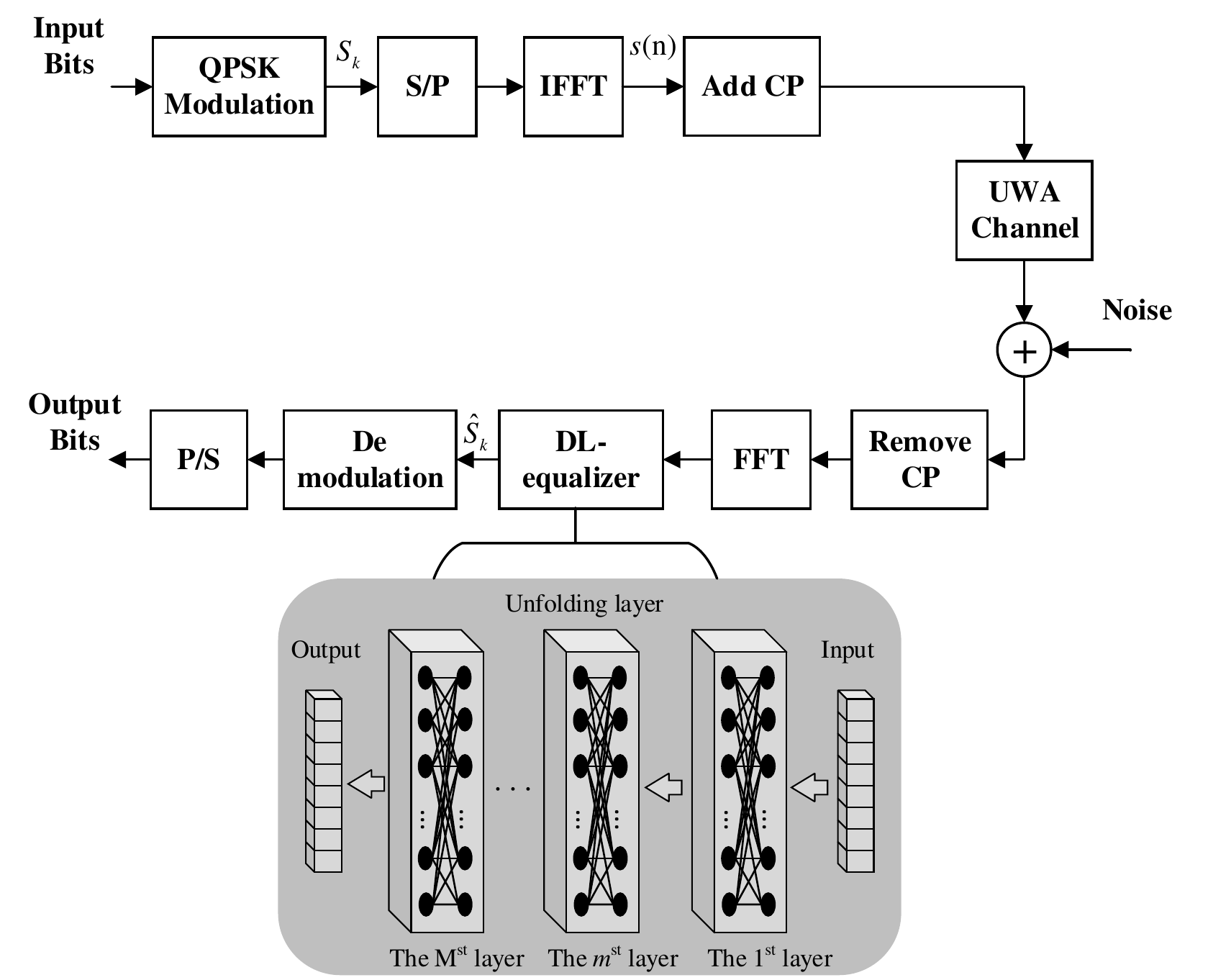}
	\caption{\textcolor{black}{UWA-OFDM system with DL based signal equalization.}}\label{framework}
\end{figure}
\textcolor{black}{
This section considers a basic single-input single-output (SISO) UWA-OFDM system. 
Fig. \ref{framework} describes the system architecture with the proposed DL-based equalizer. The DL-based equalizer employs an unfolding algorithm. The essence of the unfolding-based equalizer is to stack multi-layer neural, and each layer connection is designed according to a specific classical algorithm. 
}

At the transmitter, the input bits are mapped into a modulated symbol to obtain $S_k$ where $k$ represents the transmitted signal on the $k$-th subcarrier. Then, the transmitted symbol  $S_k$ is converted to the time domain after the $N$-point inverse fast Fourier transform (IFFT). The corresponding process can be expressed as
\begin{equation}
s(n)=\frac{1}{N} \sum_{n=0}^{N-1} S_k e^{-j \frac{2 \pi}{N} n k},
\end{equation}
where $s(n)$ means the $n$-th time sample in a OFDM symbol.

Later, $s(n)$ is arranged into blocks by a parallel-to-serial (P/S) converter, and a cyclic prefix (CP) is employed. Then, the modulated signal is converted to a passband and sent over UWA channels. In general, the length of  CP is longer than the channel maximum delay, and the ISI can be eliminated. 
\textcolor{black}{However, the doubly-selective channel  makes the OFDM symbol suffer serve ICI\cite{b24}.}
Correspondingly, the received signal $y(n)$, that is the transmitted signal $s(n)$ is transmitted over the channel $h(n,l)$, can be expressed as
\begin{equation}\label{channeltheory}
y(n)=\sum_{n=0}^{N} \sum_{l=0}^{L-1} h(n, l) s(n-l)+z(n),
\end{equation}
where $h(n,l)$ means the doubly-selective complex gain of the $l$-th path at the $n$-th moment, $L$ represents the total number of signal transmission paths, and $z(n)$ represents additive white Gaussian noise (AWGN). 

At the receiver, the CP of the received signal should be removed, and the serial-to-parallel (S/P) conversion is follows. After that, the received signal is processed by discrete Fourier transform (DFT). The vector ${\bm Y}=\mathscr{F}(\bm y)$ represents the received signal in the frequency domain.
The time-varying doubly-selective channel $\bm h$ can be converted into the frequency domain by a two-dimensional DFT (2D-DFT) $\bm {H}=2\mathscr{D}\mathscr{F}(\bm h)$.  The frequency domain channel matrix can be expressed as,
\begin{equation}
\begin{smallmatrix}
{\bm {H}}=\left[\begin{array}{ccccc}H_{00} & H_{(N-1) 1} & \cdots & H_{2(N-2)} & H_{1(N-1)} \\H_{10} & H_{01} & \cdots & H_{3(N-2)} & H_{2(N-1)} \\\vdots & \vdots & \ddots & \vdots & \vdots \\H_{(N-2) 0} & H_{(N-3) 1} & \cdots & H_{0(N-2)} & H_{(N-1)(N-1)} \\H_{(N-1) 0} & H_{(N-2) 1} & \cdots & H_{1(N-2)} & H_{0(N-1)}\end{array}\right].
\end{smallmatrix}
\end{equation}
where ${\bm H}\in\mathbb{E}^{N\times N}$  reflects the mutual interference between subcarriers.

The signal transmission process can be modeled as
\begin{equation}\label{LSorign}
{\bm Y}={\bm H} {\bm S}+{ \bm Z},
\end{equation}
where the received vector $\bm{Y}=\mathscr{F}(\bm y)$ represents the received signal in the frequency domain with ${Y}\in\mathbb{E}^{N}$, transmitted vector ${\bm S}=\mathscr{F}(\bm s)$ means the transmitted signal in the frequency domain with $\bm {S}\in \mathbb{C}^{N}$, and noise vector ${\bm Z}=\mathscr{F}(\bm z)$ means the noise in the frequency domain with size $N$. The frequency domain received symbol ${\bm Y}$ can be equalized with the DL receiver to get the predicted constellation. Then, the output bits can be gained through the demodulation and decision modules. 

\begin{figure}[t]
	\centering
	\subfigure[]{
		\begin{minipage}[]{0.5\linewidth}
			\centering
			\includegraphics[width=4cm]{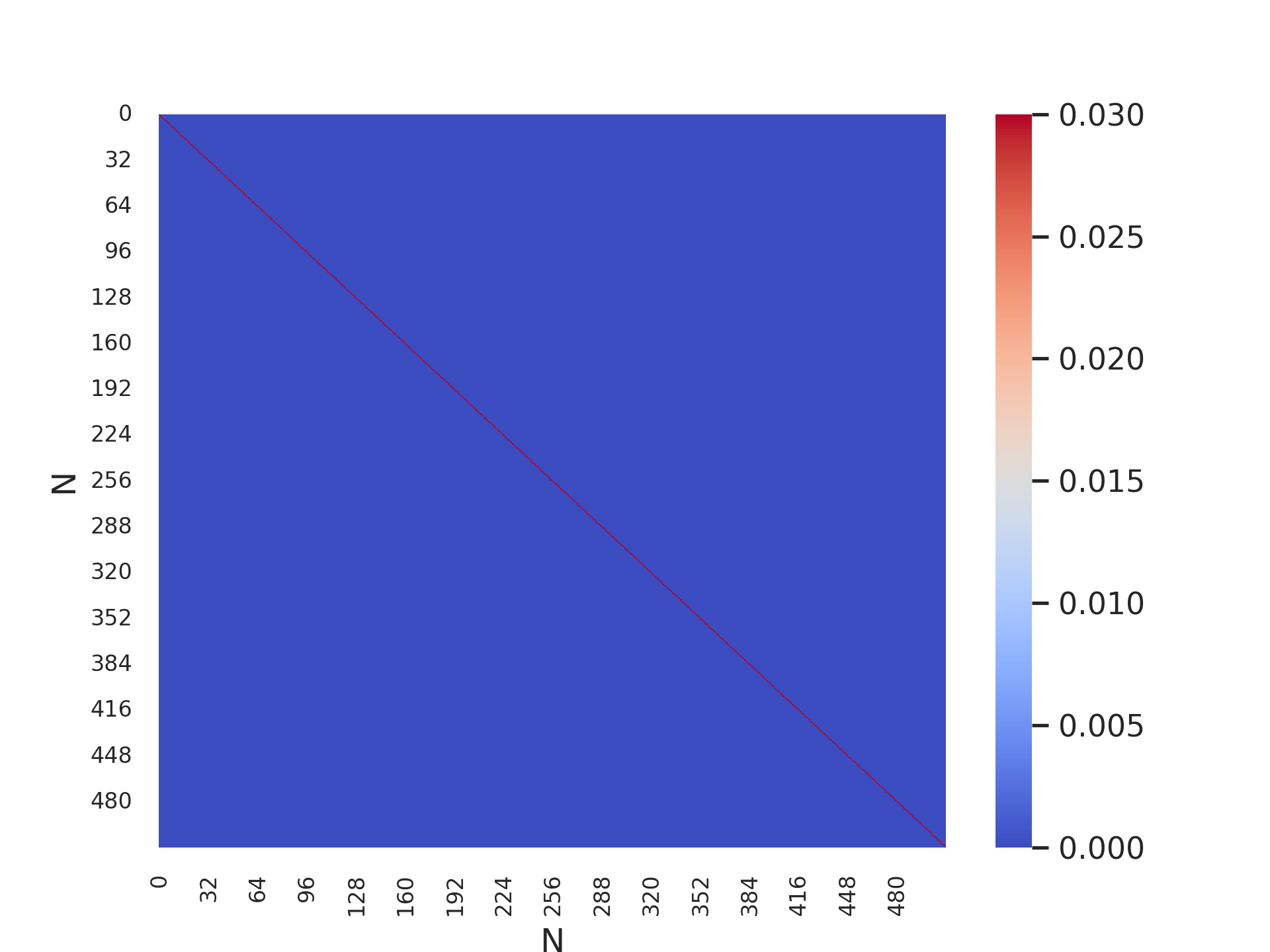}\label{heatmapppa}
		\end{minipage}%
	}%
	\subfigure[]{
		\begin{minipage}[]{0.5\linewidth}
			\centering
			\includegraphics[width=4cm]{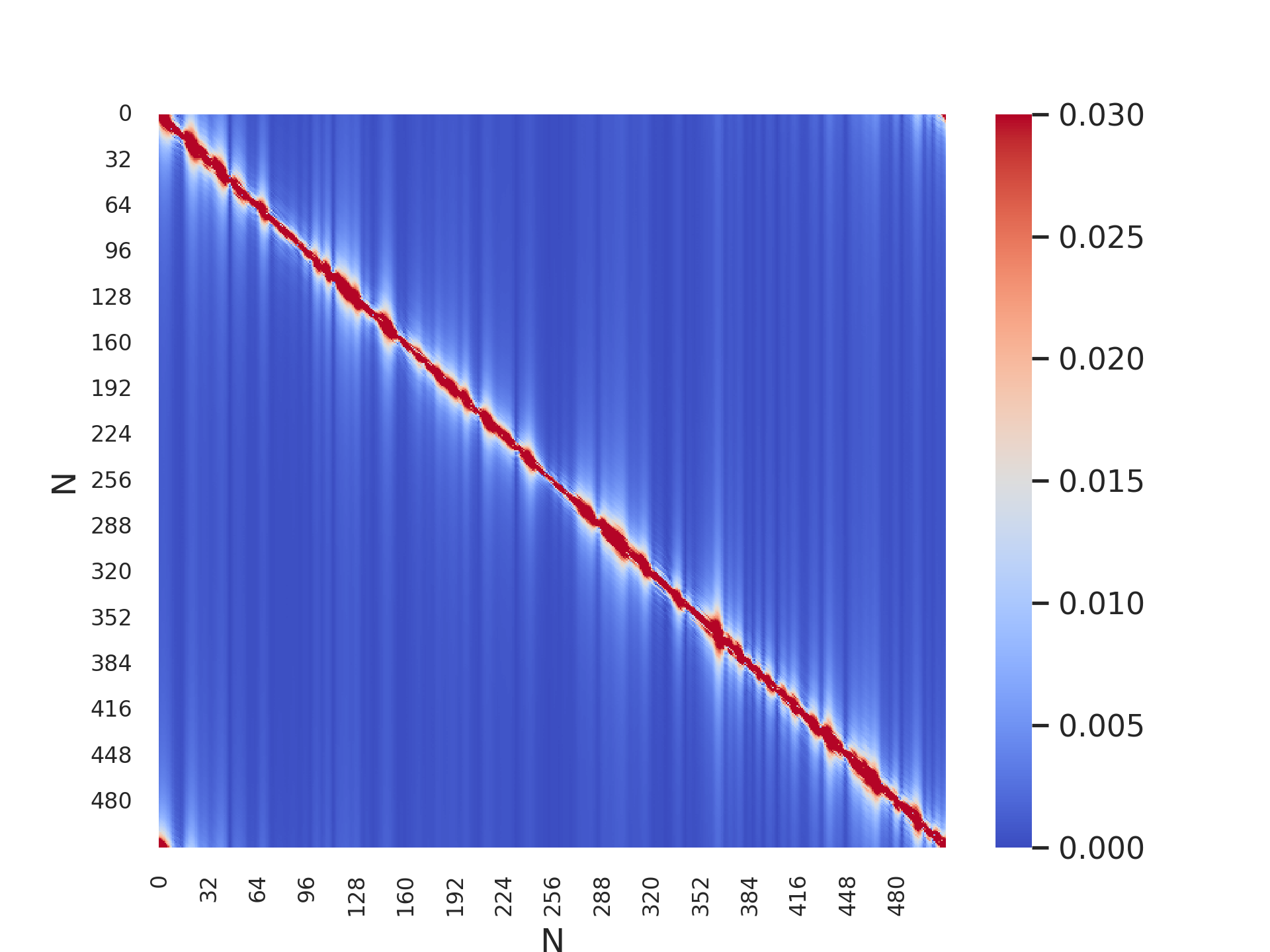}\label{heatmapppb}
		\end{minipage}%
	}%
	\centering
	\caption{Heatmap of baseband equivalent channel matrix with subcarrier number $N=512$ in the frequency domain: a) quasi-static channel heatmap b) doubly-spread channel heatmap.}\label{heatmappp}
\end{figure}

For a better understanding, we show the frequency domain channel matrix using channel heatmap in Fig. \ref{heatmappp}. 
In detail, the quasi-static channel heatmap is shown in Fig. \ref{heatmapppa}, whose energy is concentrated on the frequency matrix's diagonal. On the contrary,  for a doubly-selective channel, the energy of the channel matrix, shown in Fig. \ref{heatmapppb},  is distributed around the diagonal smoothly. The banded area is irregular but can be recognized as an approximate banded matrix.

\subsection{Imperfect CSI}\label{AA}

\textcolor{black}{
Actually, it is challenging to obtain perfect CSI, especially the underwater scenario.  We assume that imperfect CSI $\hat {\bm H}$ is used to make a recovery.
In order to evaluate the robustness of the proposed network in practice, we define a random error variable $\bm {\vartheta}$ to simulate imperfect CSI. The time domain imperfect CSI can be expressed as:}
\begin{equation}\label{imperfectE}
{ \hat {\bm{h}}} = \bm h + \bm \vartheta.
\end{equation}
\textcolor{black}{
According to (\ref{LSorign}), the ZF equalized symbol can be expressed in the frequency domain,
\begin{equation}
{ \hat {\bm{S}}}= { \hat {\bm{H}}}^{-1}{\bm Y}.
\end{equation}
The frequency domain imperfect CSI  ${\hat{\bm H}}=\mathscr{F}(\hat{\bm h})$  can be expressed as:
\begin{equation}\label{imperfectE}
{ \hat {\bm{H}}} = \bm H + \bm \varTheta,
\end{equation}
where ${{\bm \varTheta}}=\mathscr{F}(\hat{\bm \vartheta})$ means the random error in the frequency domain. 
The elements of the random error  $\bm \vartheta$ have zero  mean value, and the standard deviation of $\sigma$ generated from the normal distribution $\mathcal N(0,\sigma^2 I)$. When the CSI is perfect, $\bm \vartheta $ is set to all-zero.}

\section{UNDERWATER DEEP NETWORK}
\textcolor{black}{
	In this section, we elaborate on the UDNet equalizer framework.   In general, the deep unfolding approach is used to unfold the iteration of a specific algorithm into a single layer. Then,  we stacked $M$ layers to form a deeper equalizer.
	Fig. \ref{erframe} shows a specific layer of the UDNet.  
	The deep unfolding technique is employed to reference the MMSE algorithm for the design of DL-equalizer architecture.
	Compared with the black-box-designed method, the unfolding-based algorithms can be well explained and interpreted, which is more intuitive.
	Therefore, we utilize the unfolding approach and construct UDNet based on the MMSE structure to improve equalization performance. 
	The details of the procedure can be described into three parts, including data pre-processing, layer framework design, and optimization criteria.
}

%

\subsection{Data Pre-processing}\label{AA}
\subsubsection{Input pre-processing}
Considering the input of the general neural network must be a real number,  we first need to convert the complex signal into real form.
Concretely, the input complex signals of UDNet can be rewritten as follows,
\begin{equation}\label{preA}
\dot{\bm Y}=\left[\begin{array}{c}\mathfrak{R}({\bm Y}) \\\mathfrak{I}({\bm Y})\end{array}\right], {\dot{\bm S}}=\left[\begin{array}{c}\mathfrak{R}(\hat{\bm S}) \\\mathfrak{I}({\hat {\bm S}})\end{array}\right],
\end{equation}
\begin{equation}\label{preB}
{\dot {\bm Z}}=\left[\begin{array}{c}\mathfrak{R}({\bm Z}) \\\mathfrak{I}({\bm Z})\end{array}\right], {\dot {\bm H}}=\left[\begin{array}{cc}\mathfrak{R}({\bm H}) & -\mathfrak{I}({\bm H}) \\\mathfrak{I}({\bm H}) & \mathfrak{R}({\bm H})\end{array}\right],
\end{equation}
where  ${\dot {\bm Y}}\in\mathbb{E}^{2N}$,   ${\dot {\bm H}}\in\mathbb{E}^{2N\times2N}$,  ${\dot {\bm Z}}\in\mathbb{Z}^{2N}$, and  ${\dot {\bm S}}\in\hat{\mathbb{C}}^{2N}$ are pre-processed received signal, CSI, noise and transmitted signal in the frequency domain, respectively.

\subsubsection{Output pre-processing}
The constellation diagram is widely used in digital communications.  The classical equalizer is designed according to the MMSE criterion. 
We pre-process the form of the output data to treat the signal equalization as a classification task instead of a regression task.
We  perform one-hot encoding of the reparameterized discrete modulated signal constellation ${\mathcal{\bm C}}=\left\{\bm c_{1}, \ldots,\bm c_{i}, \ldots,   \bm c_{|{{\kappa}}|}\right\}$. 
The digital modulation then can be specified as a $\kappa$-classes problem.

For each possible $c_{i}$, we arrange a unit vector $\bm {{p}}_{i} \in \mathbb{E}^{|\hat{\mathbb{C}}|}$. We define the mapping function $\bm c=\varepsilon_{o}({\bm p})$, where $\bm c_{i}=\varepsilon_{o}\left(\bm{p}_{i}\right)$ for $i=1, \ldots,|\hat{\mathbb{C}}|$. For example,  the quadrature phase shift keying (QPSK) constellations after the one-hot encoding function can be expressed as
\begin{equation}
\begin{array}{r} 
\bm c_{1}=[1, 1] \rightarrow \bm {p}_{1}=  [1,0,0,0] \\ \bm c_{2}=[-1, 1] \rightarrow \bm {p}_{2}= [0,1,0,0] \\ \bm c_{3}=[-1, -1]  \rightarrow \bm {p}_{3}= [0,0,1,0] \\\bm c_{4}=-[-1, 1]\rightarrow \bm {p}_{4}=[0,0,0,1]\end{array}.
\end{equation}

\begin{figure}[tb!]
	\captionsetup{font={small}}
	\centering
	\includegraphics[width=7cm]{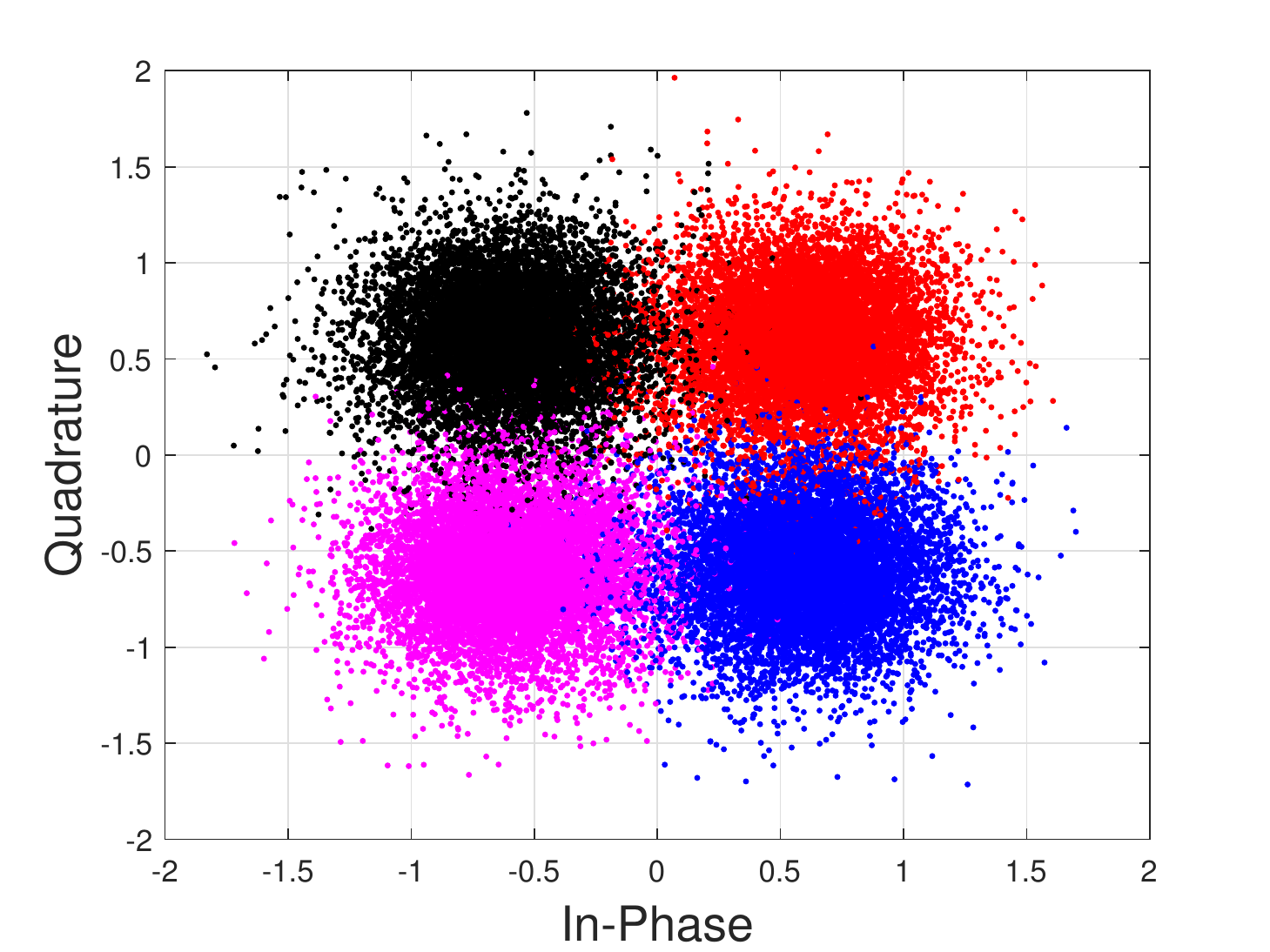}
	\caption{\textcolor{black}{Constellation diagrams of QPSK.}}\label{QPSKdiag}
\end{figure}

\textcolor{black}{
	Fig. \ref{QPSKdiag} shows the QPSK constellation diagrams. 
	The constellation diagram can  display the position and distribution of the constellation points.  For example, the QPSK is a four-category problem, intuitively. 
}
\begin{figure*}[t]
	\captionsetup{font={small}}
	\centering
	\includegraphics[width=16cm]{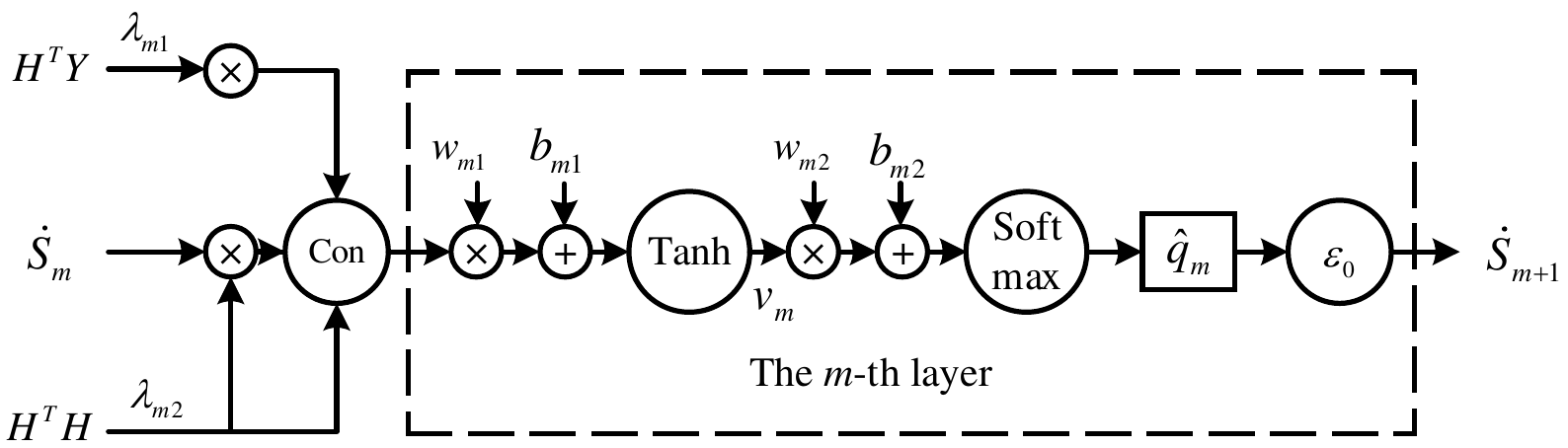}
	\caption{\textcolor{black}{A flowchart of the $m$-th layer of UDNet with classification structure.}}\label{erframe}
\end{figure*}

\subsection{Fully Connected Layer}\label{AA}
The ML equalizer involves an exhaustive search and is the optimal equalizer in the minimum joint probability of error for detecting all the symbols simultaneously\cite{b21}. 
The ML algorithm can be expressed as
\begin{equation}
\hat{{\bm S}}=\arg \underset{\bm S \in\hat{\mathbb{C}}^{2N}}{\min }\|{\bm Y}-{\bm H} {\bm S}\|.
\end{equation}
Unfortunately, the computational complexity of ML will increase exponentially with the increase of sub-carriers $N$. Especially since the sub-carriers number of OFDM  in the UWA  conditions is relatively large, the ML algorithm is highly complicated. 
Besides, ML is sensitive to channel estimation, which makes engineering applications difficult.
In practice, ML is usually replaced by sub-optimal  MMSE equalizers. Hence, we take the fundamental MMSE as a reference and unfold the MMSE to improve the equalizer performance.

According to  (\ref{LSorign}), (\ref{preA}) and (\ref{preB}),  the corresponding relationship between  pre-processed ${\hat Y}$ and ${\hat H}$ can be defined as follows,
\begin{equation}
{\dot {\bm H}}^{T} {\dot {\bm Y} }={\dot {\bm H}}^{T} {\dot {\bm H}} {\dot {\bm S}}+{\dot {\bm H}}^{T} {\dot {\bm Z}}. 
\end{equation}

\textcolor{black}{
The flowchart of the $m$-th layer of UDNet is shown in Fig. \ref{erframe}.
Further, the gradient steepest descent method calculates the global minimum point. Therefore, the unfolding iterative algorithm of the network can be formulated as
\begin{equation}
{\bm v_m} = {\rho_{{\rm{Tanh}}}}\left( {{\bm w_{m1}}\left[ {\begin{array}{*{20}{c}}
		{{\bm S_m}}\\
		{{\lambda _{m1}}{\bm H^T}\bm Y}\\
		{{\lambda _{m2}}{\bm H^T}\bm H{\rm{ }}{\bm S_m}}
		\end{array}} \right] + {\bm b_{m1}}} \right),
\end{equation}
\begin{equation}
{ {\hat{\bm q}_{m}}} = {\rho_{{\rm{Softmax}}}}\left( {{\bm w_{m2}}{\bm v_m} + {\bm b_{m2}}} \right),
\end{equation}
where  $\bm v_{m}$ a temporary variable of each layer, $\dot {\bm S}_{m}$ is the $m$-th iteration value, ${\bm W}_{m}$ means the weight matrix of the $m$-th layer, ${\bm n}_{m}$ means the bias term of the $m$-th layer, and $\lambda_{m 1}$ and $\lambda_{m 2}$ represent the step size parameters of the $m$-th layer, respectively.
Meanwhile, the initial value of  $\dot{\bm {S}}_{0}$  is assigned by ZF equalization to increase the convergence rate during training.
We employ two nonlinear functions for the $m$-th layer to eliminate the channel distortion.
Because of the universal approximation theorem~\cite{WNT}, any two nonlinear functions can fit any function.  
The learnable parameters of the $m$-th layer is $\bm \Xi_m = \{\bm w_{m1} ,\bm b_{m1}  , \bm w_{m2} , \bm b_{m2}\}$. The whole network  learnable parameters are included in $\bm \Psi = \{\bm \Xi_1 , \dots, \bm \Xi_m, \dots,\bm  \Xi_M\}$.
The first activation function is  ${\rho_{{\rm{Tanh}}}}$ with parameters $ \{w_{m1} , n_{m1}\}$  in order to map the output within $[-1, 1]$, which is consistent with the constellation. The second activation function is  ${\rho_{{\rm{Softmax}}}}$ with parameters $ \{w_{m2} , n_{m2}\}$, which 
maps the estimated symbol into one-hot format. Then, the QPSK de-mapping function $\varepsilon$ is used, which converts one-hot coding to the QPSK symbol to ensure the input of the following layer is still constellation information. 
Via a mapping function, we obtain $\hat{{\bm S}}_{m+1}=\varepsilon_{o}(\hat{{\bm q}}_{m})$. Then,  $\hat{{\bm S}}_{m+1}$ is used as the input of the $(m+1)$-th layer.
}

\begin{figure}[t!]
	\captionsetup{font={small}}
	\centering 
	\includegraphics[width=9cm]{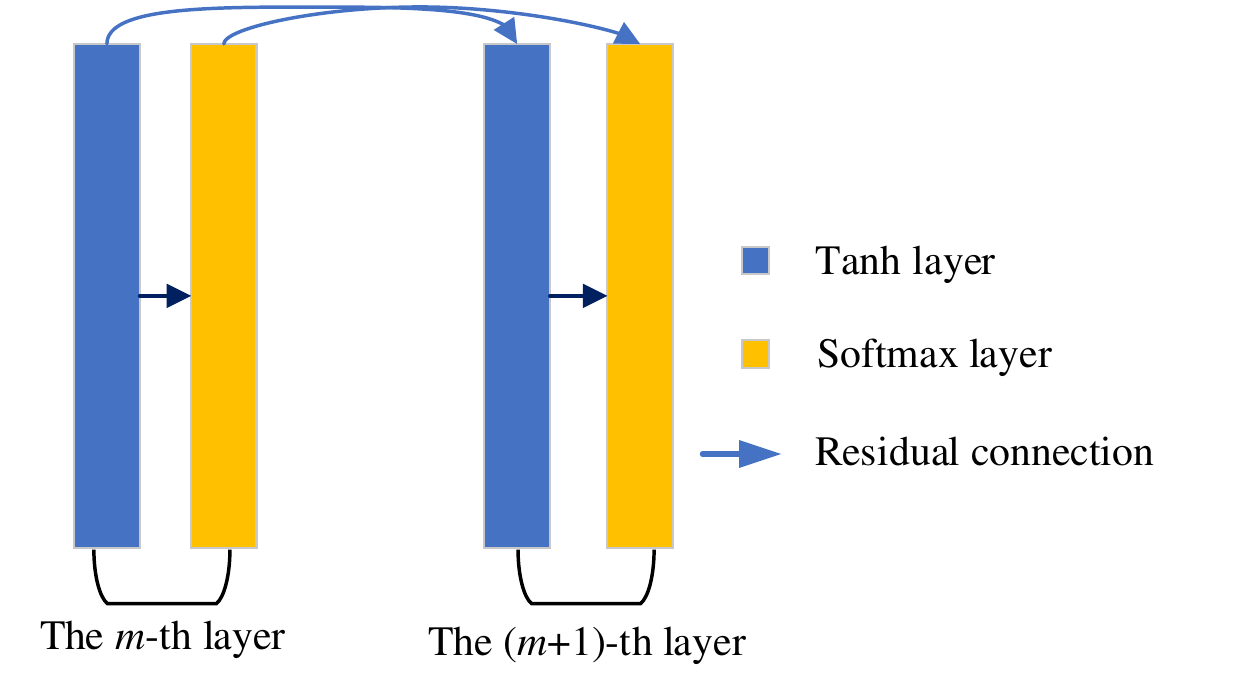}
	\caption{\textcolor{black}{The structure of residual connection.}}\label{LAYERS}
\end{figure}

Moreover, we employ cross-layer residual connections to avoid the proposed network's vanishing gradient problem. The specific framework is shown in Fig. \ref{LAYERS}.
\textcolor{black}{Specifically, we connect the $m$-th and the $(m+1)$-th output of $Tanh$. In the same way, we also connect the $m$-th and the $(m+1)$-th output of $Softmax$. The cross-layer residual connections solve the vanishing gradient problem and accelerate the training procedure. } 


\subsection{KL Criterion}\label{AA}
\textcolor{black}{We use classification ideas to optimize the equalizer. The equalizer based on the MMSE is not achieving the minimum symbol error rate (SER) performance. Unlike the minimum MSE, we pursue the distribution of the prediction constellation diagram closest to the true distribution. In this way, the SER performance can be improved.
Hence, the minimum KL criterion is employed to evaluate the gap between the true label and estimated predictions. We perform one-hot encoding with a constellation diagram to utilize the position and distribution information. The QPSK constellation points are regarded as the label, and the corresponding estimated symbol is the prediction value.
The KL divergence is a nonsymmetric measure of the difference between two probability distributions $P$, and $Q$. 
The loss function can be expressed as:}

\textcolor{black}{
\begin{equation}
\begin{array}{l}
{D_{KL}}(\bm S;\hat {\bm S}\left( {\bm Y,\bm H;\bm \Psi )} \right)\\
= \sum\limits_{m = 1}^M {\sum\limits_{k = 1}^N {{P_m}({S_k})\log \frac{{{P_m}({S_k})}}{{{Q_m}({{\hat S}_k})}}} } \\
= \sum\limits_{m = 1}^M {\sum\limits_{n = 1}^N {({P_m}({S_k})\log ({P_m}({S_k})) - {P_m}({S_k})\log ({Q_m}({{\hat S}_k})))} } \\
= \sum\limits_{m = 1}^M {(H({P_m}(\bm S)) - H({P_m}(S),{Q_m}(\hat {\bm S})))} \\
= - \sum\limits_{m = 1}^M {H({P_m}(\bm S),{Q_m}(\hat {\bm S}))} ,
\end{array}
\end{equation}
where  $P_m{(S)}$ represents the true distribution. The measured $Q_m({\hat S})$ typically represents a predicted distribution.
${D_{KL}}(\bm S;\hat {\bm S}\left( {\bm Y,\bm H; \bm \Phi)} \right)$ denotes the KL divergence, where $\Phi$ represents the training parameters in the network,   ${\bm Y}$ and ${\bm H}$ are the input of the network.  
}

\textcolor{black}{
Moreover, in the training stage, the information entropy of a given dataset is a fixed constant, and the derivative of the constant is zero. Therefore, cross-entropy can be directly used to train a model. That is to say, minimizing KL divergence is equivalent to minimizing cross entropy, and the calculation of cross entropy is more straightforward.}
\subsection{Sliding Structure of OFDM Signal}
\begin{figure*}[t]
	\centering
	\subfigure[NOF]{
		\begin{minipage}[t]{0.33\linewidth}
			\centering
			\includegraphics[width=6cm]{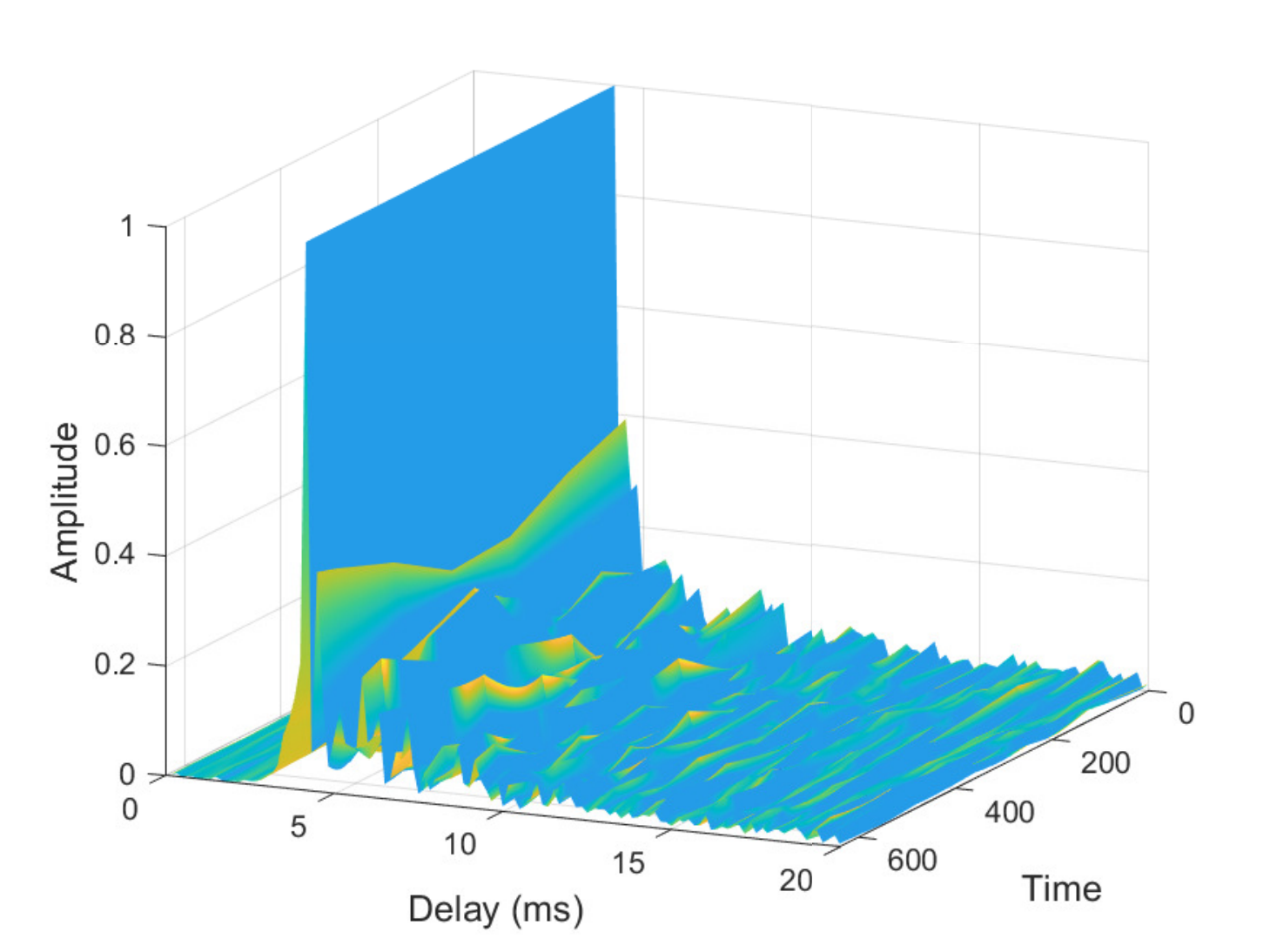}
		\end{minipage}%
	}%
	\subfigure[NCS]{
		\begin{minipage}[t]{0.33\linewidth}
			\centering
			\includegraphics[width=6cm]{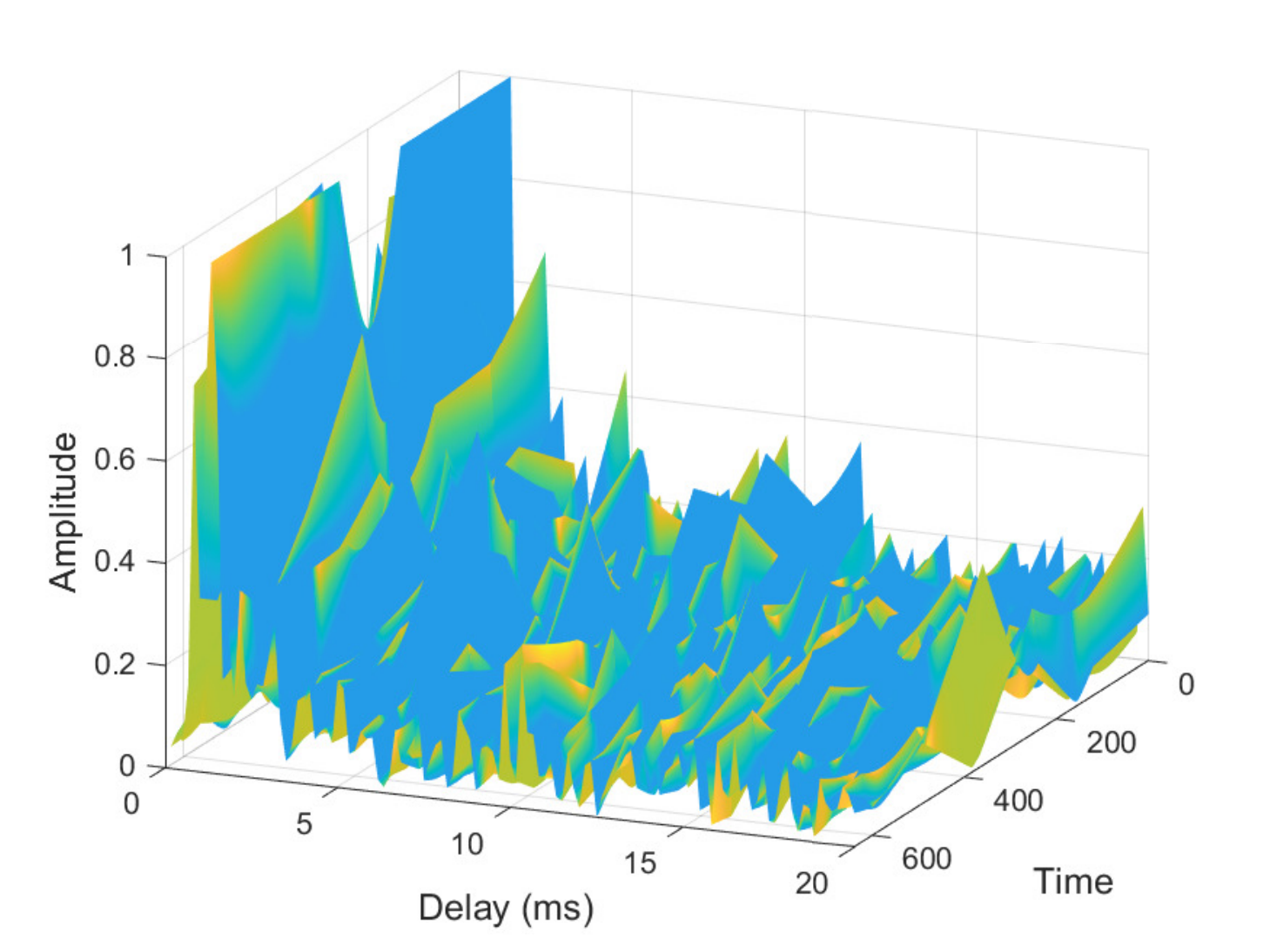}
		\end{minipage}%
	}%
	\subfigure[SIM]{
		\begin{minipage}[t]{0.33\linewidth}
			\centering
			\includegraphics[width=6cm]{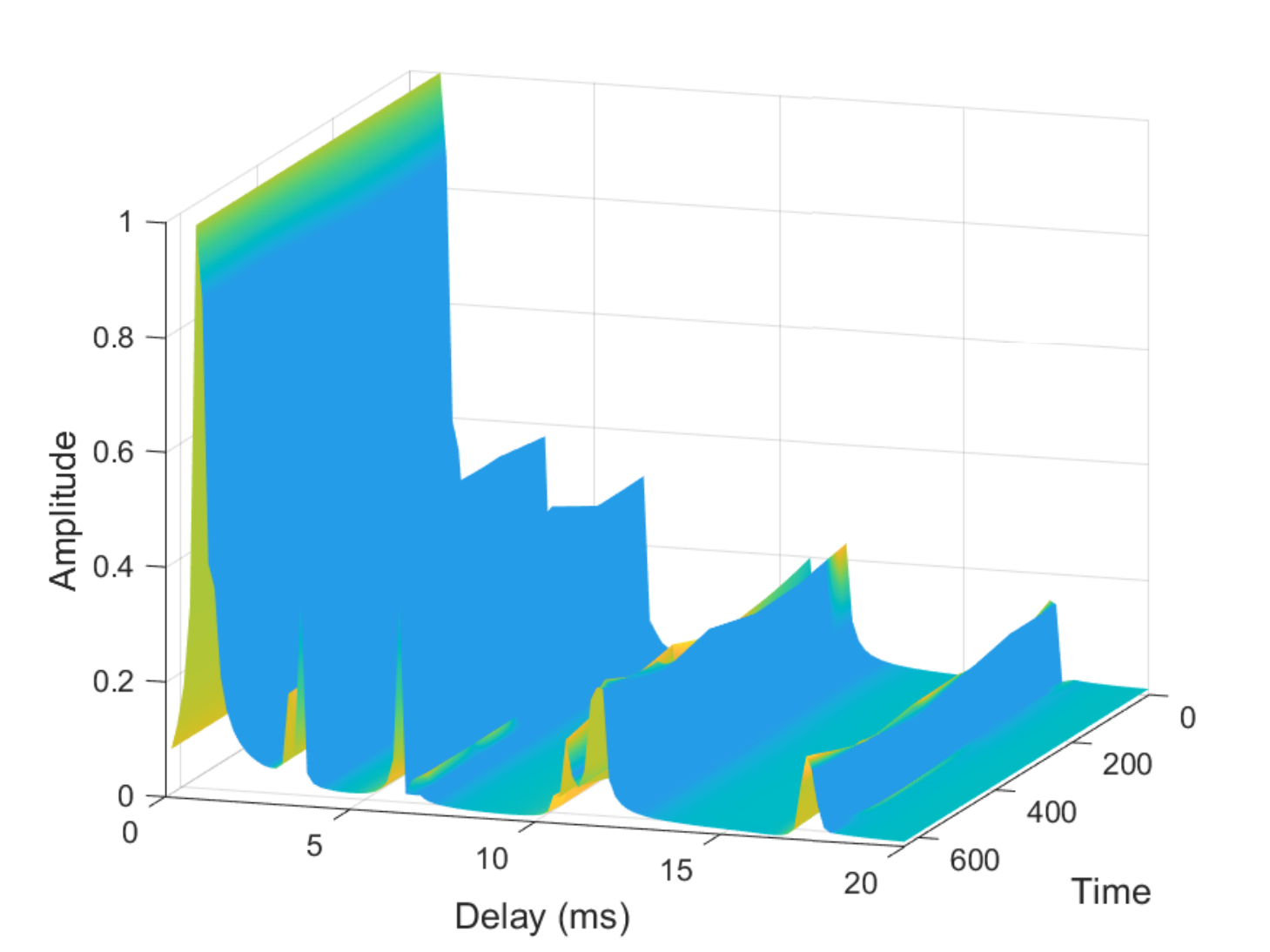}
		\end{minipage}
	}%
	\centering
	\caption{\textcolor{black}{Time-varying impulse response of three  kinds of UWA scenarios.}}\label{CHANNEL}
\end{figure*}

Let the matrix ${H}$ be split into sliding blocks represented as
\begin{equation}
{\bm H}=\left[\begin{array}{llll}{\bm \phi_{1}} & & & \\& {\bm \phi_{2}} & & \\& & \ddots & \\& & & {\bm \phi_{J}}\end{array}\right],
\end{equation}
where $J$ means the number of sliding blocks, and ${\bm \phi_{j}}$ represents the $j$-th sliding block of ${\bm H}$. Meanwhile, the received signal ${\bm y}$, the estimated value of the transmitted signal $\hat{{\bm S}}$ can be expressed as
\begin{equation} 
{\bm Y}=\left[\begin{array}{llll}{\bm \Gamma_{1}} & {\bm \Gamma_{2}} & \cdots & {\bm \Gamma_{J}}\end{array}\right],
\end{equation}

\begin{equation} 
\hat{{\bm S}}=\left[\begin{array}{llll}{\bm \Upsilon_{1}} & {\bm \Upsilon_{2}} & \cdots & {\bm \Upsilon_{J}}\end{array}\right],
\end{equation}
where $\bm \Gamma_{j}$ and $\bm \Upsilon_{j}$ represents the $j$-th sliding block of ${\bm Y}$ and $\hat{{\bm S}}$, respectively. Block-like segmented sliding processing is performed on the input signals ${\bm H}$, ${\bm Y}$, and $\hat{{\bm S}}$, thereby reducing the problem of insufficient computing resources and memory. Therefore, we get the network output with sliding blocks as
\begin{equation} 
\hat{{q}}=\left[\begin{array}{llll}{\bm \Omega_{1} }& \bm \Omega_{2} & \cdots & \bm \Omega_{J}\end{array}\right],
\end{equation}
where $\bm \Omega_{j}$ represents the $j$-th sliding block of $\hat{{q}}$.


\begin{table}[t]
	\begin{center}
		\centering
		\caption{\textcolor{black}{Summary of System Parameters}}\label{tab1}   
		\begin{tabular}{ccc}  
			\toprule   
			Parameters & Value \\
			\midrule   
			Modulations & QPSK\\
			Subcarriers & 512/1,024/2,048 \\
			Channel length & 120 taps\\
			Sliding size & 32*32\\
			Training SNR &  25dB \\
			Testing SNR &   10/15/20/25/30/35dB\\
			Optimizer & Adam\\
			Layer number & 22\\
			Learning rate & 0.001\\
			Batch size & 2,000\\
			Epochs & 200\\
			CIRs & 20,000\\
			\bottomrule   
		\end{tabular}
	\end{center}
\end{table}

\section{Experimental RESULTS}

\subsection{Training Parameter Settings}\label{AA}
In this section, we introduce the detailed parameters of the network. We utilize TensorFlow architecture and python3.6 for training. Table~\ref{tab1} lists the system parameters of UDNet training. \textcolor{black}{Meanwhile, the network is trained using the adaptive moment estimation (Adam) gradient descent method. In detail, the total layer number of UDNet is set to $M=22$. The initial learning rate is set to 0.001, and the training batch is set to $2,000$. The total training epoch is $200$ for each epoch. Further, the number of CIRs we used in the training procedure is $45 \times 6000$.} The channel length is set to 120 taps. The channel matrix is divided into $32 \times  32$ size of sliding block and input to the UDNet. In the training stage, the SNR is fixed at 25dB.
It should be noted that we explored the comparison experiment of complexity on NVIDIA GeForce RTX 3090.

\begin{table}[t]
	\begin{center}
		\centering
		\caption{\textcolor{black}{Parameters of Test Conditions}}
		\label{tab2}   
		\begin{tabular}{cccc}  
			\toprule   
			Parameters & NOF& NCS & SIM\\
			\midrule   
			Environment & Fjord & Shelf & Default \\
			Range&  750m & 540m &500m-8,000m \\
			Water depth & 10m & 80m & 100m \\
			Transmitter depl & Bottom & Bottom & Suspended\\
			Receiver depl & Bottom & Bottom & Suspended\\
			Doppler coverage & 7.8Hz & 31.4Hz & Uncalculated\\
			Type & Measured& Measured & Simulated\\
			Quantity&\textcolor{black}{216,180} &	\textcolor{black}{15,480}&	\textcolor{black}{61,440}\\
			\bottomrule   
		\end{tabular}
	\end{center}
\end{table}

The number of sub-carriers of the selected OFDM signal is $N$=512 by default. Due to the large sub-carrier of the UWA signal and the computer memory limitation, the algorithms we compared are all processed with sliding structures. The signal modulation method is QPSK. When we compare the performance of different algorithms, we will use the following naming convention:\\
\textbf{MMSE:} The classical MMSE algorithm of equalizers.\\
\textbf{DFE:} The decision feedback equalizer (DFE) algorithm.\\
\textbf{SDR:} A decoder based on semidefinite relaxation using efficient interior point solver in\cite{b31}.\\
\textbf{DetNet:} The detector architecture based DL proposed in \cite{b21},  but modified to render it suitable for time-varying channel.\\
\textbf{SCN:} Proposed method in \cite{b13}, which is a novel equalizer method, combining deep unfolding and sliding structure.\\
\textbf{UDNet:} The equalizer network architecture proposed in this paper.

\subsection{Training and Testing Dataset}\label{AA}
Table~\ref{tab2} shows the parameters of the UWA doubly-selective channel impulse response (CIR) dataset. Our dataset contains simulated and measured UWA CIRs, which are used in short-distance and shallow-water communication scenarios. Fig. \ref{CHANNEL} shows the CIRs of three underwater channels. The raw CIRs were measured at Norway—Oslofjord (NOF) and Norway—Continental Shelf (NCS)\cite{b32}. Moreover, we created a part of data by BELLHOP (SIM-B), which is a well-known UWA channel simulation method\cite{b34}. The dataset generated by NOF of 75\% is used for offline training of UDNet, and the remaining 25\% is employed for online testing.
\textcolor{black}{ It is worth mentioning that the UDN only needs to train once over the NOF channel, and it can work over NCS and SIM channels, which shows its generalization. Besides, other CIRs measured under different environments are considered for online testing. The corresponding dataset can be founded in \cite{b33}.}

\begin{figure}[tb!]
	\captionsetup{font={small}}
	\centering 
	\includegraphics[width=9cm]{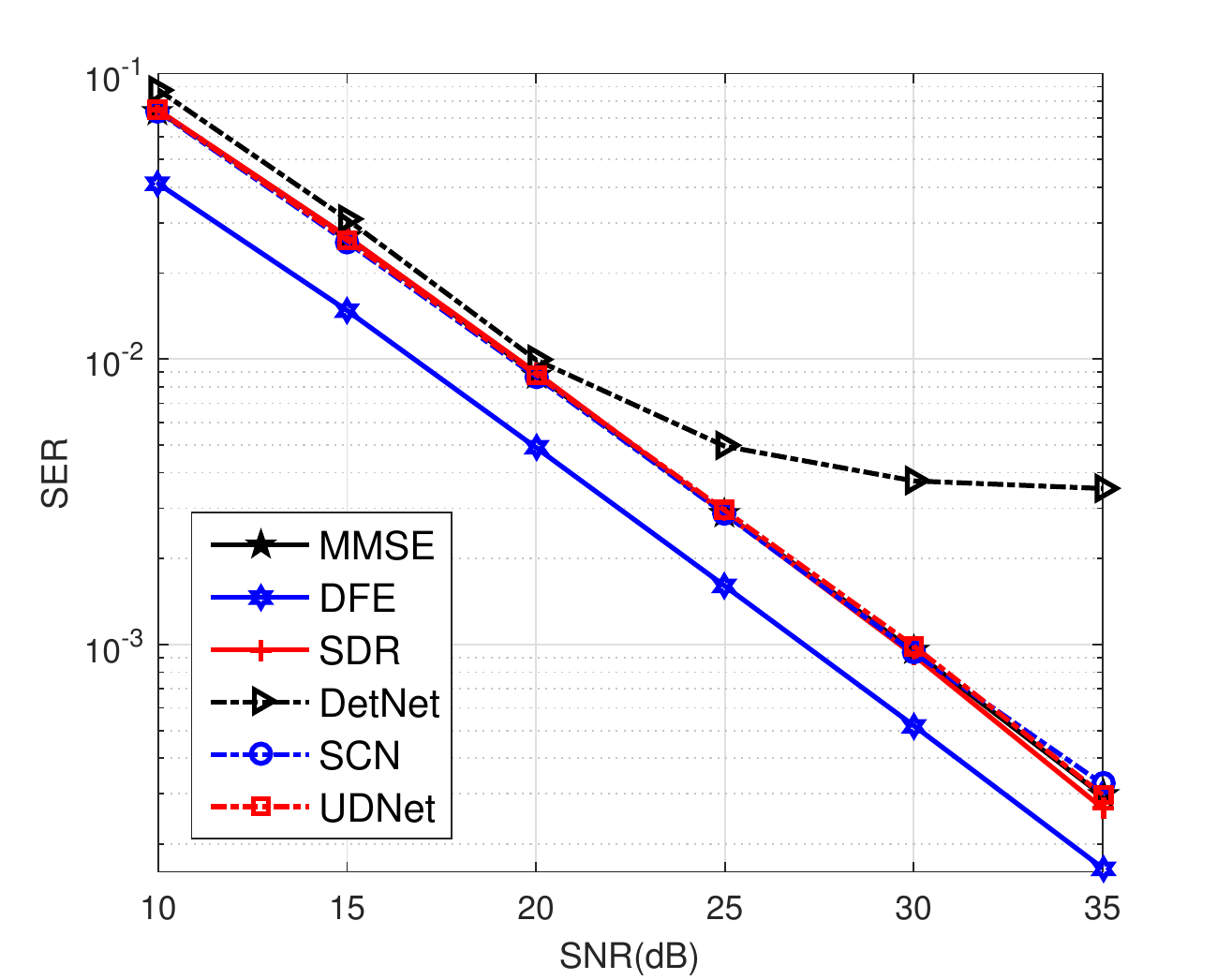}
	\caption{ \textcolor{black}{SER curves of UDNet and competing methods over the quasi-static NOF channel.}}\label{Fig7_quasiCH}
\end{figure}

\begin{figure}[tb]
\captionsetup{font={small}}
\centering 
\includegraphics[width=9cm]{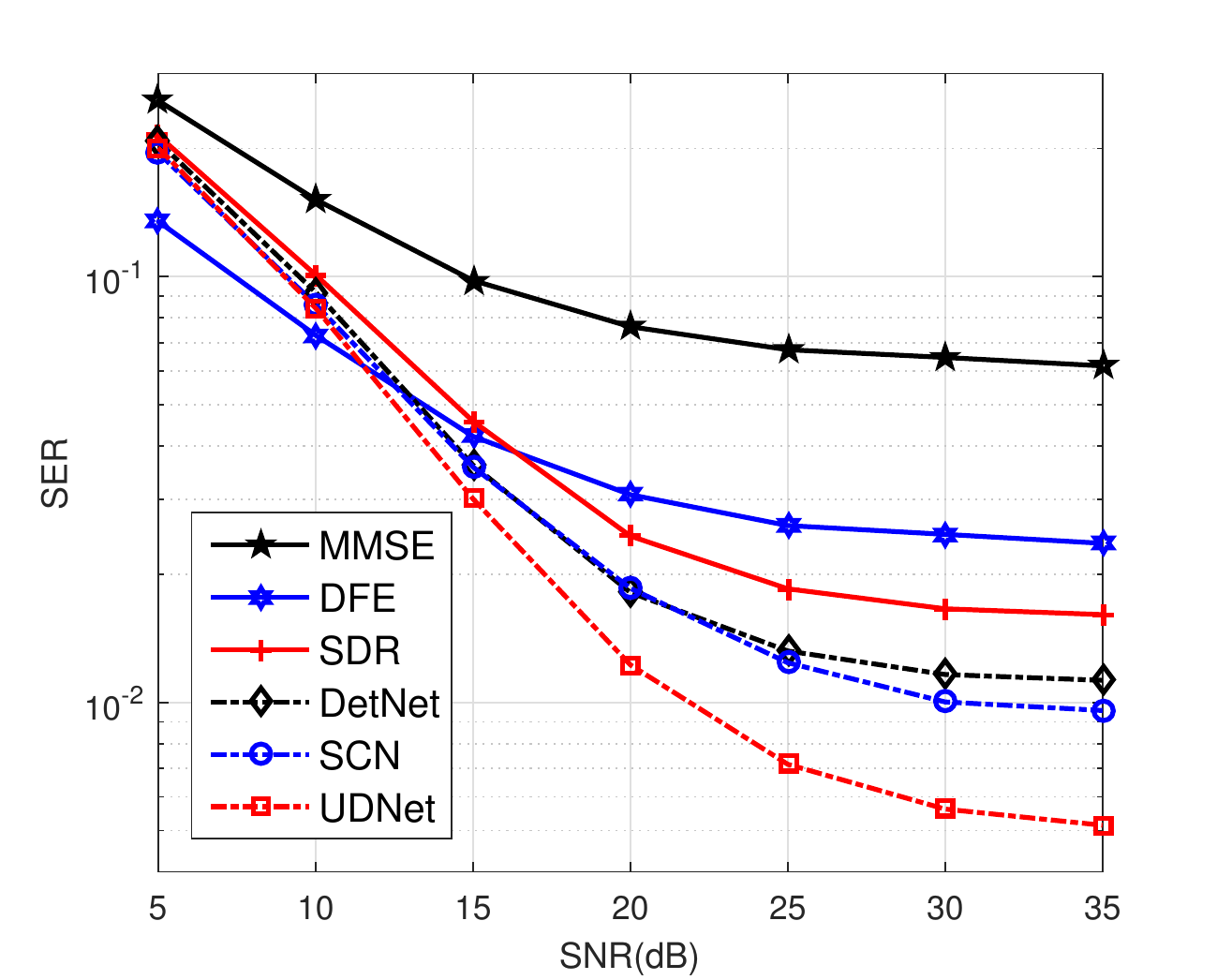}
\caption{ \textcolor{black}{SER curves  of UDNet and competing methods over the time-varying NOF channel.}}\label{Fig8_NOFqpsk}
\end{figure}
\begin{figure}[tb]
\captionsetup{font={small}}
\centering 
\includegraphics[width=9cm]{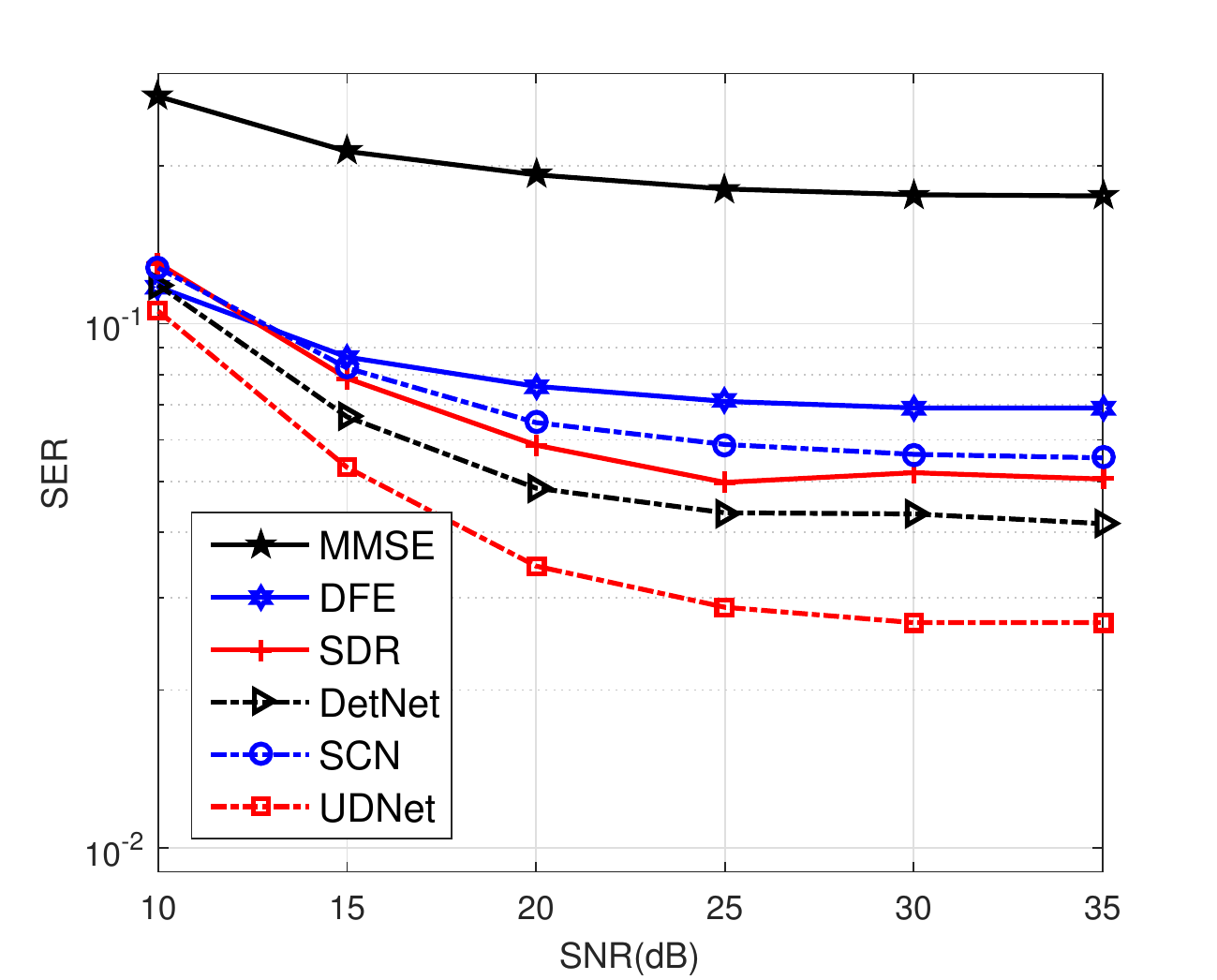}
\caption{ \textcolor{black}{SER curves  of UDNet and competing methods over the time-varying NCS channel.}}\label{Fig9_NCSqpsk}
\end{figure}
\begin{figure}[tb]
\captionsetup{font={small}}
\centering 
\includegraphics[width=9cm]{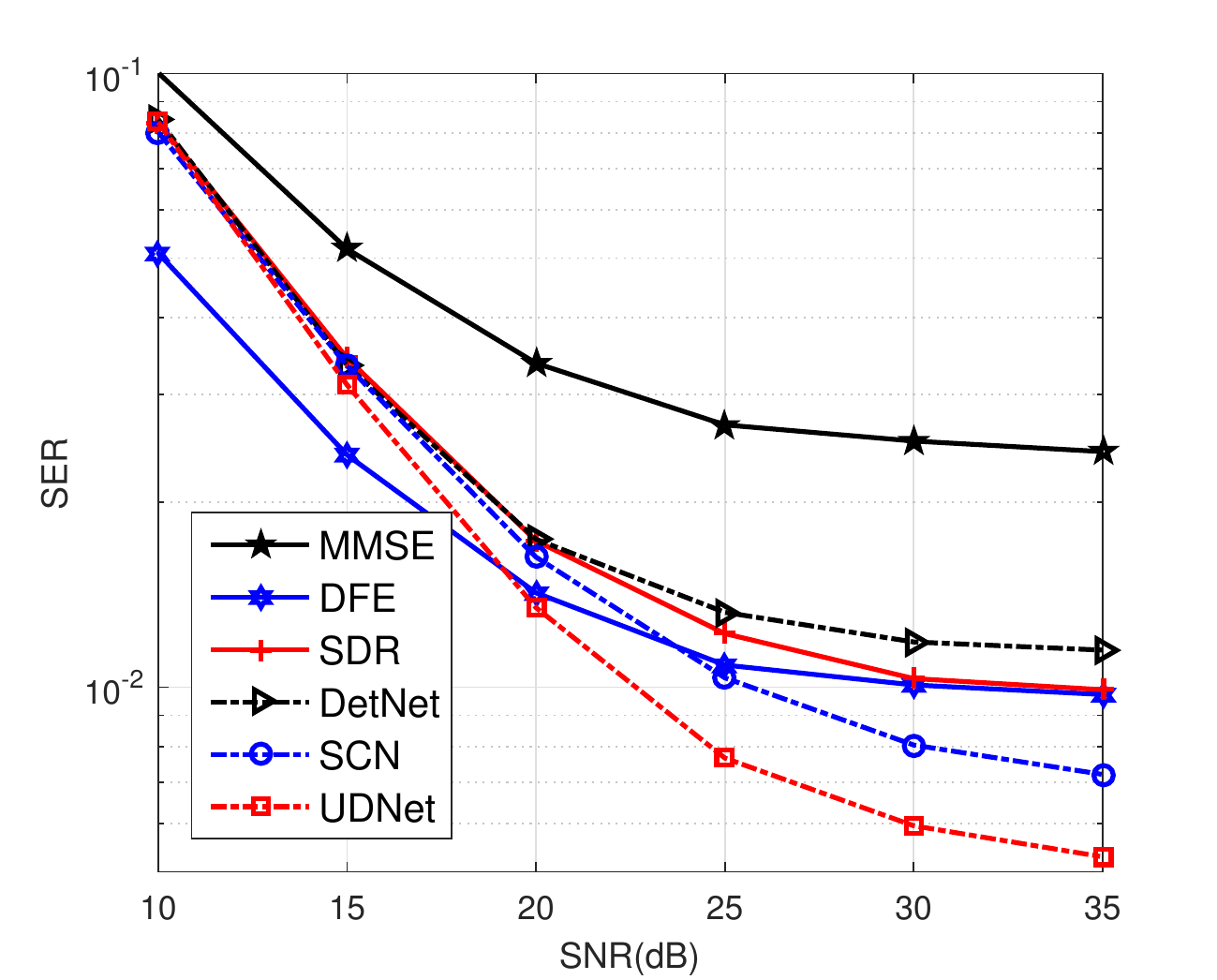}
\caption{\textcolor{black}{SER curves  of UDNet and competing methods over the time-varying SIM channel.}}\label{Fig10_SIMqpsk}
\end{figure}

\subsection{SER Performance under Different Conditions}\label{AA}
We are very concerned about the SER performance of the proposed equalizer. 
In the following experiments of different conditions,  we test the mild quasi-static conditions and the tough doubly-selective conditions with perfect CSI. 

\textit{1) Quasi-Static Channel:} In the quasi-static scenario, the UWA channel is constant within an OFDM symbol. In this case, (\ref{channeltheory}) can be simplified as
\begin{equation}
y(n)=\sum_{n=0}^{N} \sum_{l=0}^{L-1} h(l) s(n-l)+z(n),
\end{equation}
where the channel gain $h(l)$, and $z(n)$ AWGN have nothing to do with the time. 
Meanwhile, frequency domain channel matrix ${\bm H}$ only has a non-zero value on the diagonal. The ICI influence caused by the Doppler effect of ${\bm H}$ can be ignored. \textcolor{black}{Fig. \ref{Fig7_quasiCH} shows the SER curves of UDNet and competing methods over the quasi-static NOF channel. We can find that the DetNet is not suitable for quasi-static channels, and the DFE performs the best SER performance. Further, the SCN and the proposed UDNet perform similarly to the classical approaches, i.e., MMSE and SDR. The sliding approach increases the amount of training data with the same number of channels.
	Overall, the advantages of complex networks are not highlighted in simple quasi-state scenarios, but classical methods perform well. More notably, the UWA channel is rarely quasi-static.}

\textit{2) Doubly-selective Channel:} 
In general, the channel varies within an OFDM symbol. Hence, we pay attention to the doubly-selective scenario, also dubbed time-varying.
Fig. \ref{Fig8_NOFqpsk} shows SER curves of UDNet and competing methods over the time-varying NOF channel. 
It can be seen that the performance of UDNet is better than other methods between a wide range of SNR. However, when SNR is high, the downward trend of the UDNet curve slows down, and the wrong platform appears. It is due to when the sliding is performed. We ignore some non-zero values. The channel matrix is regarded as a quasi-band, and we lost part of CSI. It results in some performance penalties under high SNR.

\begin{figure}[tbp]
	\captionsetup{font={small}}
	\centering 
	\includegraphics[width=9cm]{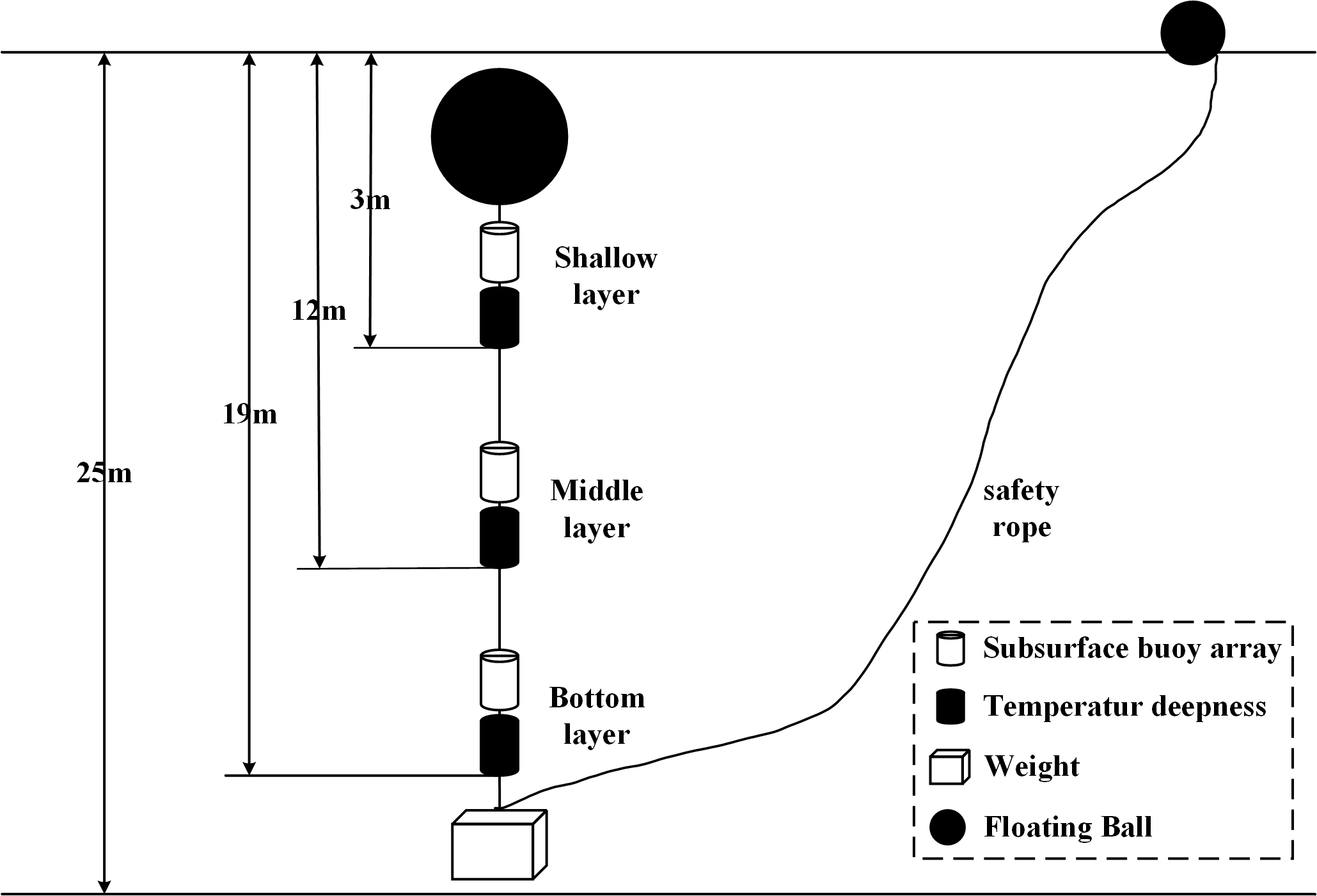}
	\caption{Bottom-mounted submersible offshore noise acquisition system off  Nanpeng island, Yangjiang, China.}\label{fig14}
\end{figure}

\subsection{Generalizations of Different Equalizers}
\textcolor{black}{
	The scene of the UWA channel is diverse, and the DL method is sensitive to data changes. The generalization is highly concerned. We compare the experimental over the other two kinds of channels, and it is marked  NCS channel is the most complicated. From Fig. \ref{Fig9_NCSqpsk} and Fig. \ref{Fig10_SIMqpsk}, we show the generalization of UDNet and competing equalizers over different channels, i.e., NCS and SIM. The DL-equalizers, i.e., DetNet, SCN, and UDN, are all trained over NOF channels.
	They can perform well over NCS and SIM. It illustrates that unfolding-based equalizers have strong generalizations. Specifically, under a challenging condition, that is, NCS channels,  UDNet consistently outperforms other algorithms in the SNR range of $0-35$dB.
	UDNet performs better than other algorithms in high SNR for the mild SIM channel, but its performance is not as good as the traditional DFE in the SNR range of $0-20$dB.
	Further,  we use a comprehensive view to compare Fig. \ref{Fig9_NCSqpsk} and Fig.  \ref{Fig10_SIMqpsk}. We can find that the SER performance ranking of equalizers is unstable. In Fig. \ref{Fig9_NCSqpsk},  the ranking of SER from the smallest to the largest in the SNR range of $20-30$dB is  UDNet$<$DetNet$<$SDR$<$SCN$<$DFE$<$MMSE. Relatively, In Fig. \ref{Fig10_SIMqpsk},  the ranking of SER in the SNR range of $20-30$dB is  UDNet$<$SCN$<$DFE$<$SDR$<$DetNet$<$MMSE. It is because the proposed UDNet employs the KL criterion to design an equalizer. The KL criterion maintains the outstanding performance and stability of UDNet.
	Simultaneously, it is elaborated that UDNet shows superior performance over complex channels, and the other algorithms have a weaker ability to deal with complex channels.}


\begin{figure}[tbp]
	\captionsetup{font={small}}
	\centering 
	\includegraphics[width=9cm]{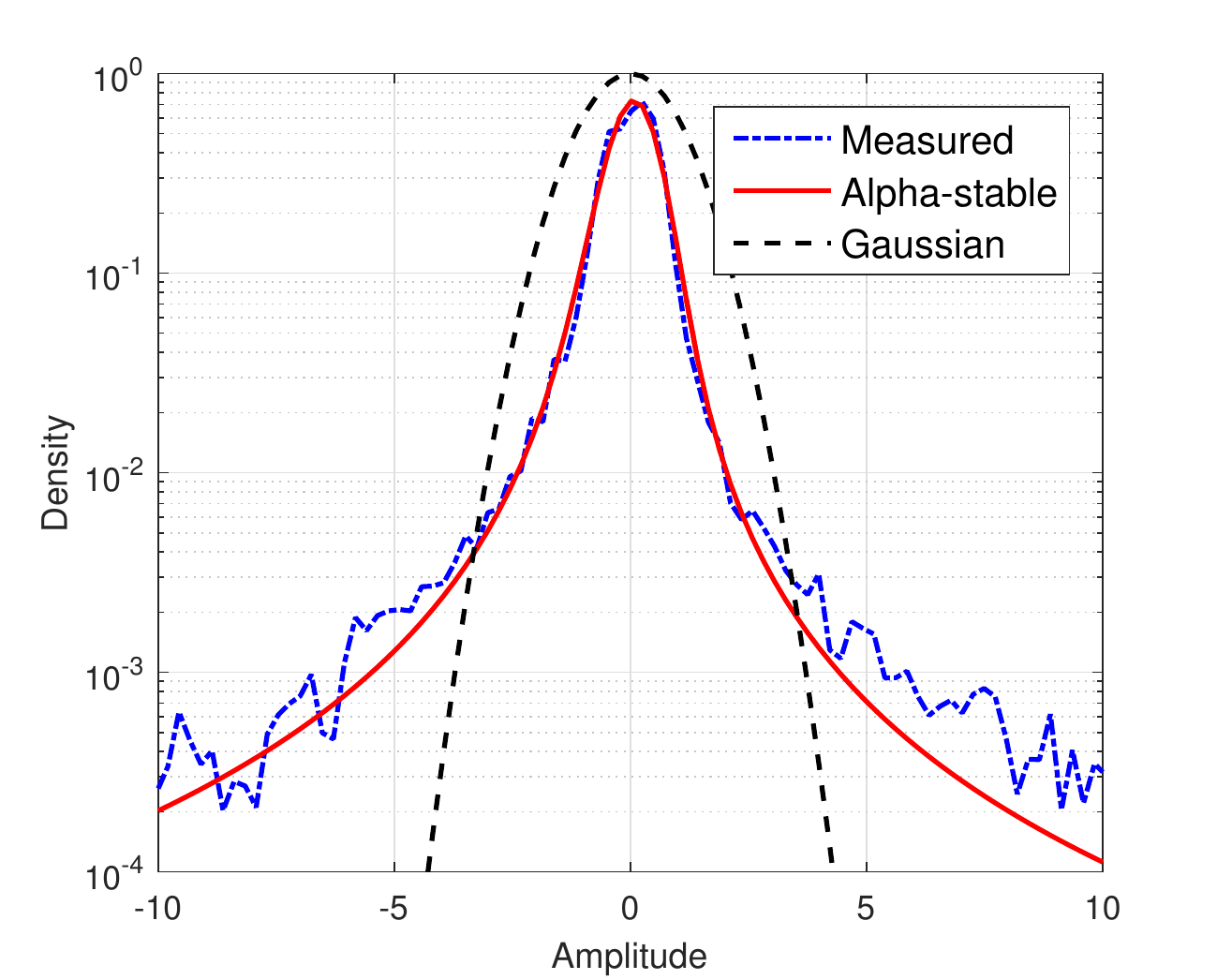}
	\caption{\textcolor{black}{Comparison of the collected noise, alpha noise and Guassian noise.}}
	\label{Noise}
\end{figure}
\begin{figure}[tb]
	\captionsetup{font={small}}
	\centering 
	\includegraphics[width=9cm]{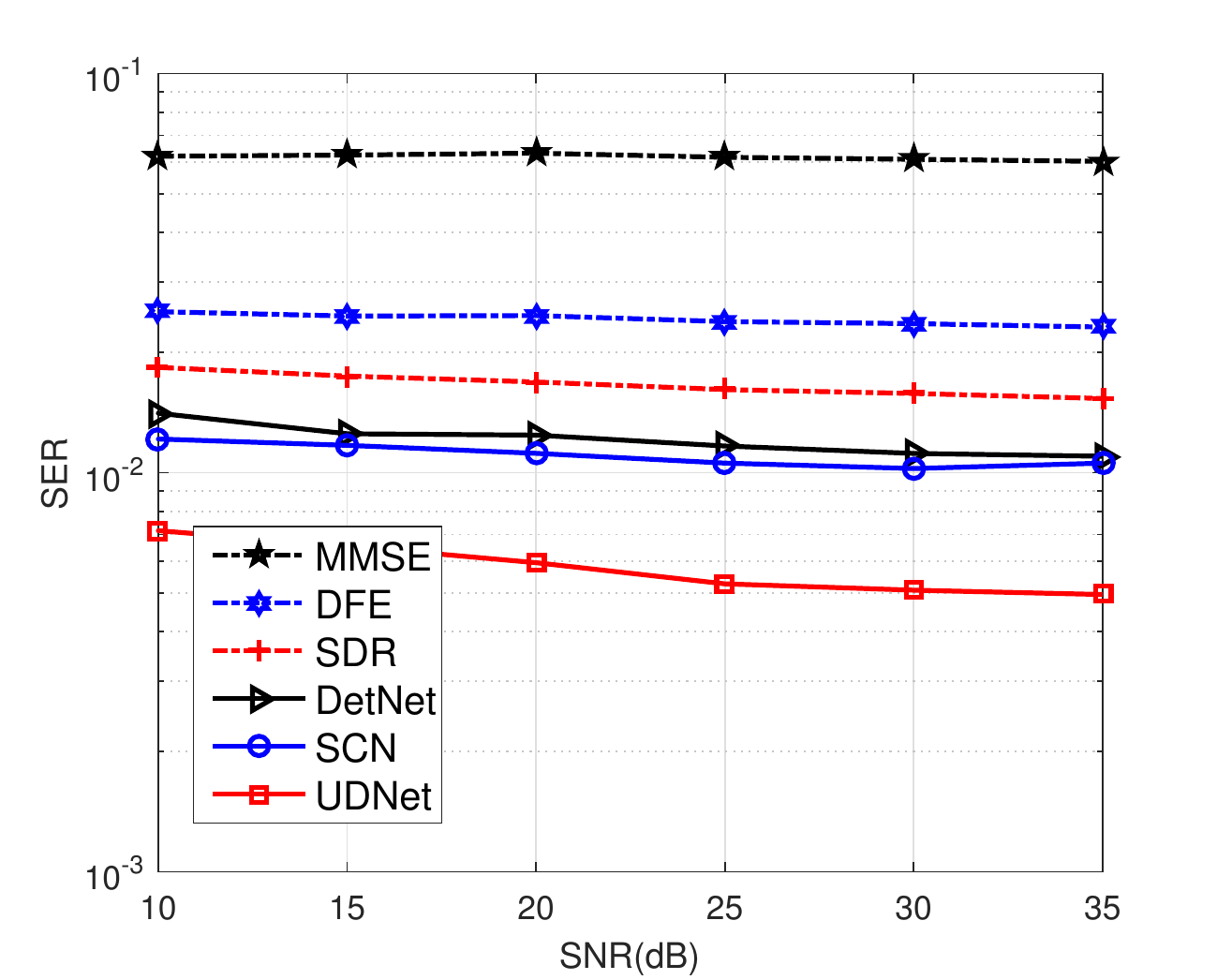}
	\caption{\textcolor{black}{Comparison of anti-noise performance of different algorithms over NOF channel.}}\label{Fig16_antiNoise}
\end{figure}

\begin{figure*}[t]
	\centering
	\subfigure[]{
		\begin{minipage}[]{0.5\linewidth}
			\centering
			\includegraphics[width=9cm]{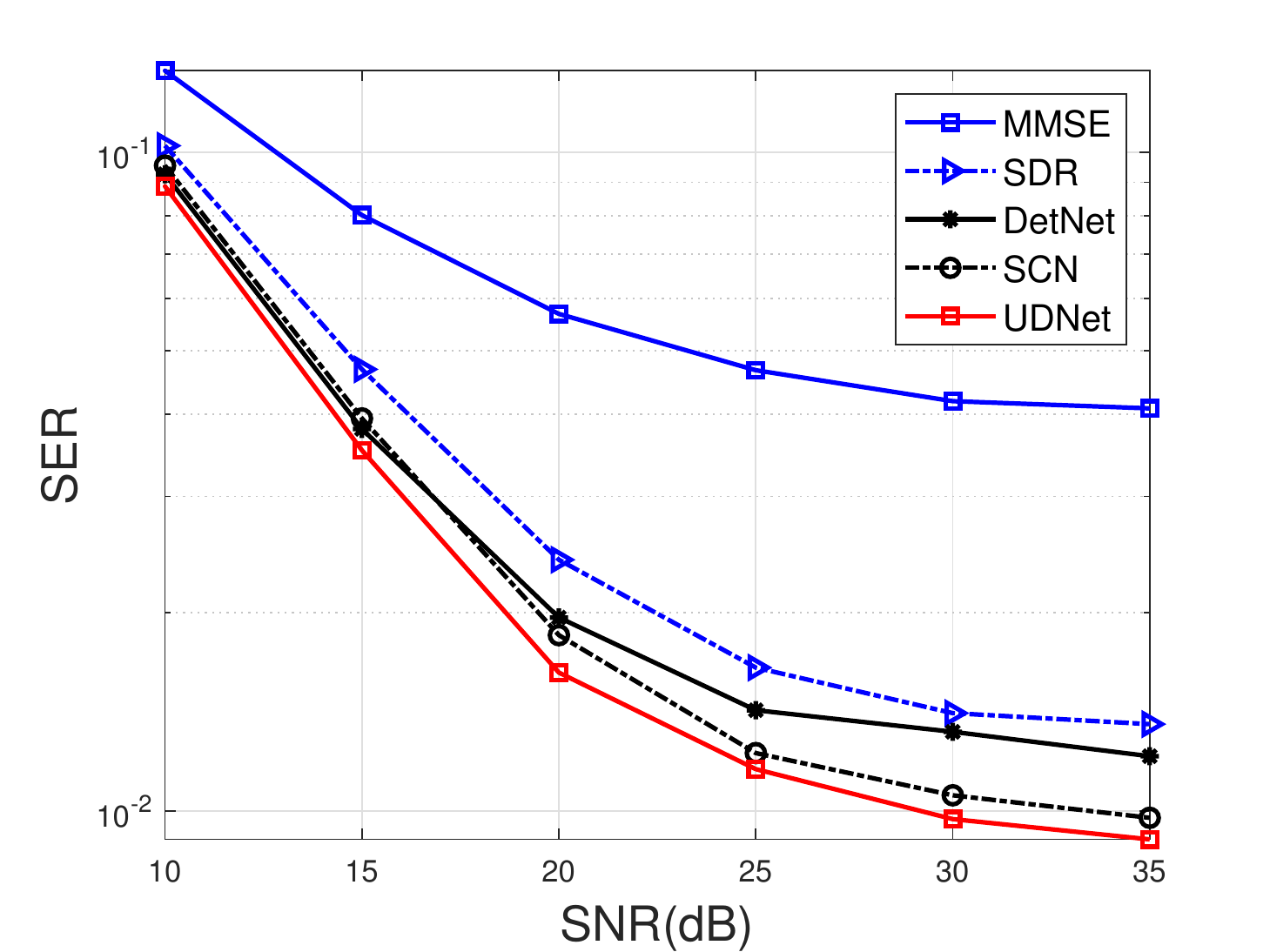}
		\end{minipage}%
	}%
	\subfigure[]{
		\begin{minipage}[]{0.5\linewidth}
			\centering
			\includegraphics[width=9cm]{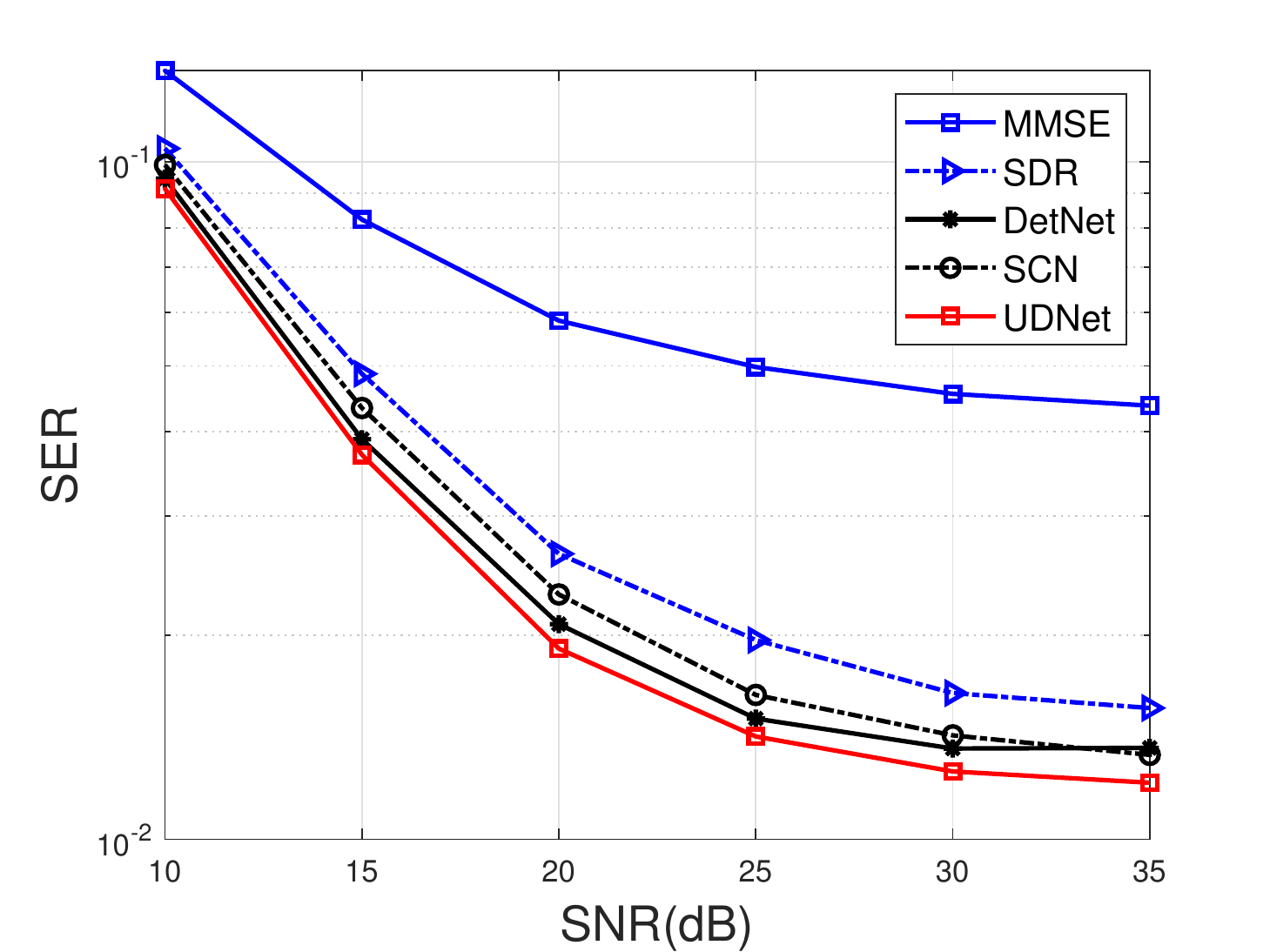}
		\end{minipage}%
	}%
	\centering
	\caption{\textcolor{black}{SER curves  of UDNet and competing equalizers over the time-varying NOF channel with imperfect CSI: a) $\sigma=0.003$. b) $\sigma=0.005$.}}\label{Fig13a_imperfect005}
\end{figure*}
\subsection{SER Performance under Collected Offshore Background Noise }\label{AA}

In this part, we first present a measured offshore background noise to provide the experiment more realistic. The marine noise of the UWA channel is a non-white Gaussian noise composed of sounds from turbulence, shipping, wind and waves, and thermal noise.   Fig. \ref{fig14} shows the noise collecting system, which adopts a bottom-mounted diving subsurface buoy. We used a self-contained acoustic recorder to collect offshore background noise, and the collection experiment lasted for two days in an offshore wind farm at  Nanpeng Island,  Yangjiang, China.  
In detail, the subsurface buoy arrays are placed in the shallow layer $3$ meters below the water surface, the middle layer is located $12$ meters below the water surface, and the bottom layer is installed $15$ meters below the water surface, respectively. The floating ball keeps the subsurface buoy suspended, and a weight is added to the bottom of the device to keep the whole device vertical. The collected noise is filtered by a band-pass filter to obtain the processed noise in the frequency  between $10$ and $18$ kHz.

Fig. \ref{Noise} shows the measured noise probability density distributions. Meanwhile, to descript the type of noise, we add Gaussian and alpha-stable Noise as benchmarks.
It is evident that the measured noise is a non-zero mean value, and its probability distribution curve is more slender than the AWGN distribution curve. It goes to the alpha-stable distribution. It is because there is a lot of impulse noise during recording.
Moreover, we use the measured noise as the background noise of the UWA signal for the anti-noise experiment of UDNet. Fig. \ref{Fig16_antiNoise} shows the experimental result of the anti-noise performance of different algorithms over the NOF channel. We can see that the SER performance of UDNet is better than other algorithms in the range of 0-35dB SNR. It clearly shows that UDNet has a more stable anti-noise performance. Concretely, it not only can deal with the additive white gaussian noise but also the measured impulse noise.
In summary,  the proposed UDNet is more robust to fix the truly offshore background noise. Moreover, the result further illustrates the advantages of UDNet in the underwater environment.

\subsection{SER Performance under Imperfect CSI}\label{AA}
\textcolor{black}{
Actually, it is challenging to acquire perfect CSI in the  UWA communication system because of the limited bandwidth resources. Therefore, to get closer to the real conditions, we validate the robustness of these equalization algorithms with imperfect CSI. 
Fig.  \ref{Fig13a_imperfect005} shows SER curves of UDNet and competing equalizers over the time-varying NOF channel with imperfect CSI. According to (\ref{imperfectE}), we add error $e$ to the perfect CSI ${\bm h}$ to simulate the imperfect CSI ${\hat {\bm h}}$. The mean value of  $\bm \vartheta$ is zero. Furthermore, the  $\sigma$  of error is set to $0.001$ and $0.003$, respectively. The $\sigma$ controls the inaccuracy degree of ${\bm h}$. The bigger $\sigma$ is, the more blurry $\hat {\bm h}$ is.
It is shown that as  $\hat {\bm h}$ deviates more, CSI is more imperfect, and the performance of all equalizer algorithms decreases. However, UDNet still performs better than other methods in the varying procedure of imperfect CSI. It is because traditional algorithms rely more on correct channel estimation, but imperfect CSI will affect its accuracy. Furthermore, the DL-based methods have better robustness and accuracy, especially when they utilize probability as the output of soft information, i.e., UDNet.}

\subsection{Impact of Clipping and Filtering Distortion}\label{AA}


\begin{figure}[tb]
	\captionsetup{font={small}}
	\centering 
	\includegraphics[width=9cm]{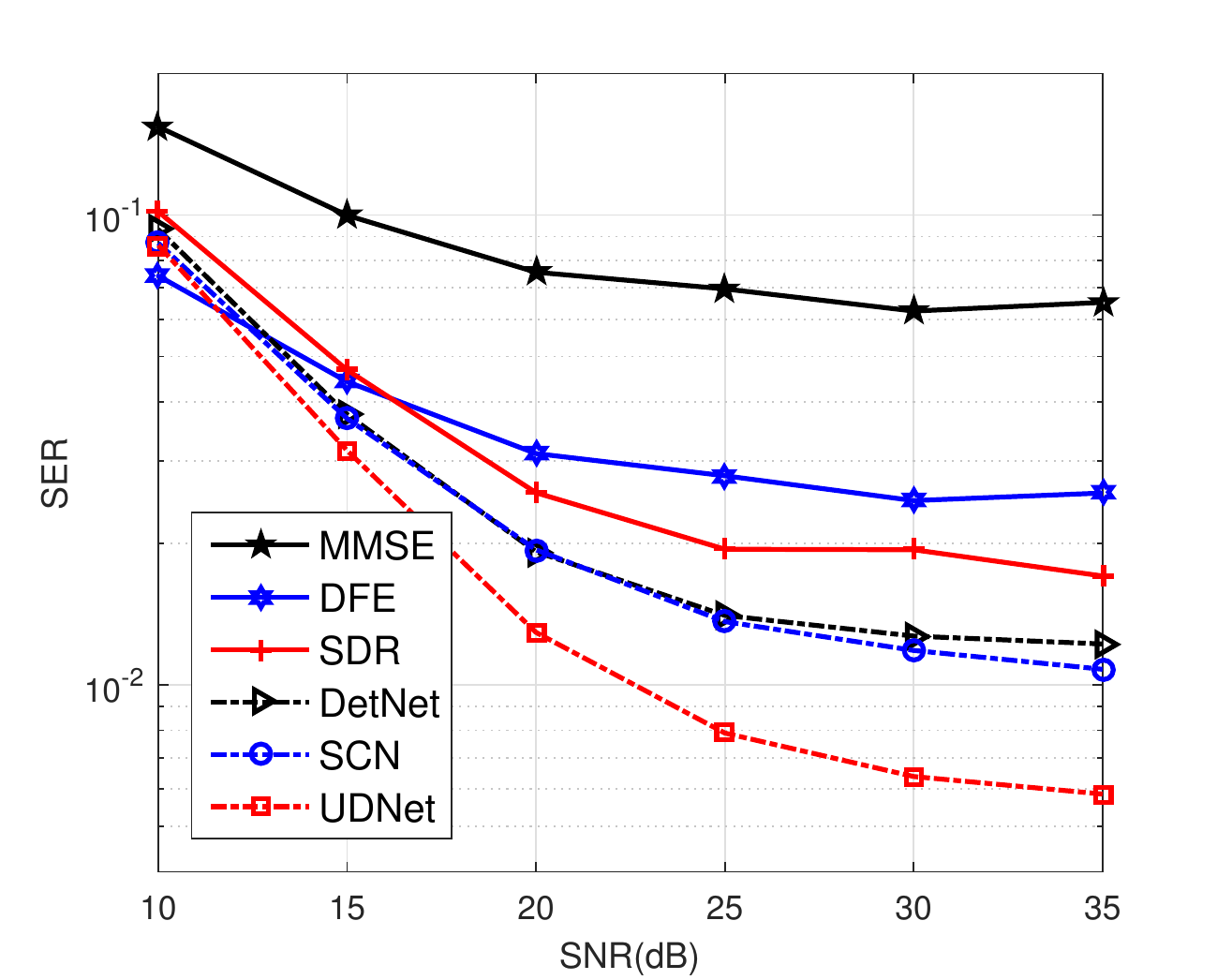}
	\caption{\textcolor{black}{Comparison of different algorithms under clipping over NOF channels.}}\label{Fig12_NOFclipping}
\end{figure}

Fig. \ref{Fig12_NOFclipping} compares UDNet and other algorithms under clipping and filtering distortion. In the UWA OFDM system, there are random interference factors, such as the attenuation of information energy and waveform distortion caused by the system's nonlinearity  channel\cite{b25}.   Therefore,  a significant shortcoming of OFDM,  the high peak-to-average power ratio (PAPR), is fatal for the UWA OFDM communications because of clipping.
The clipping will introduce nonlinear noise, and the received data may be badly distorted.  Concretely, the clipped signal can be expressed as,
\begin{equation}
c(n)=\left\{\begin{array}{ll}
c(n), & \text { if }|c(n)| \leq A \\
A e^{j \phi(n)}, & \text { otherwise }
\end{array}\right.,
\end{equation}
where $A$ is the limiting threshold, $c(n)$ is the received signal, and $\phi(n)$ is the phase of $c(n)$. 

We set the clipping ratio to the OFDM signal as $1$.  As we can observe, all methods have a decay with SER performance under the clipping noise setting. Simultaneously, it almost covers $10-35$dB SNR. It is shown that the UDNet method outperforms the other methods. This result proves that the UDNet method is more robust with the nonlinear clipping noise.

\subsection{Complexity Analysis}\label{AA}

\begin{table}[t]
	\begin{center}
		\centering
		\caption{\textcolor{black}{Complexity Comparison of DL-Based Equalizers}}
		\label{tab3}   
		\begin{tabular}{ccccc}  
			\toprule   
			Name & Flops& Paras & Memory & Multiplications and Additions\\
			\midrule   
			DetNet & 0.96M & 0.47M & 2.1MB & 22,752\\
			SCN & 0.39M & 0.20M & 0.85MB & 9,344\\
			UDNet & 0.65M & 0.33M & 1.5MB & 15,520\\
			\bottomrule   
		\end{tabular}
	\end{center}
\end{table}

\begin{figure}[tb]
	\captionsetup{font={small}}
	\centering 
	\includegraphics[width=9cm]{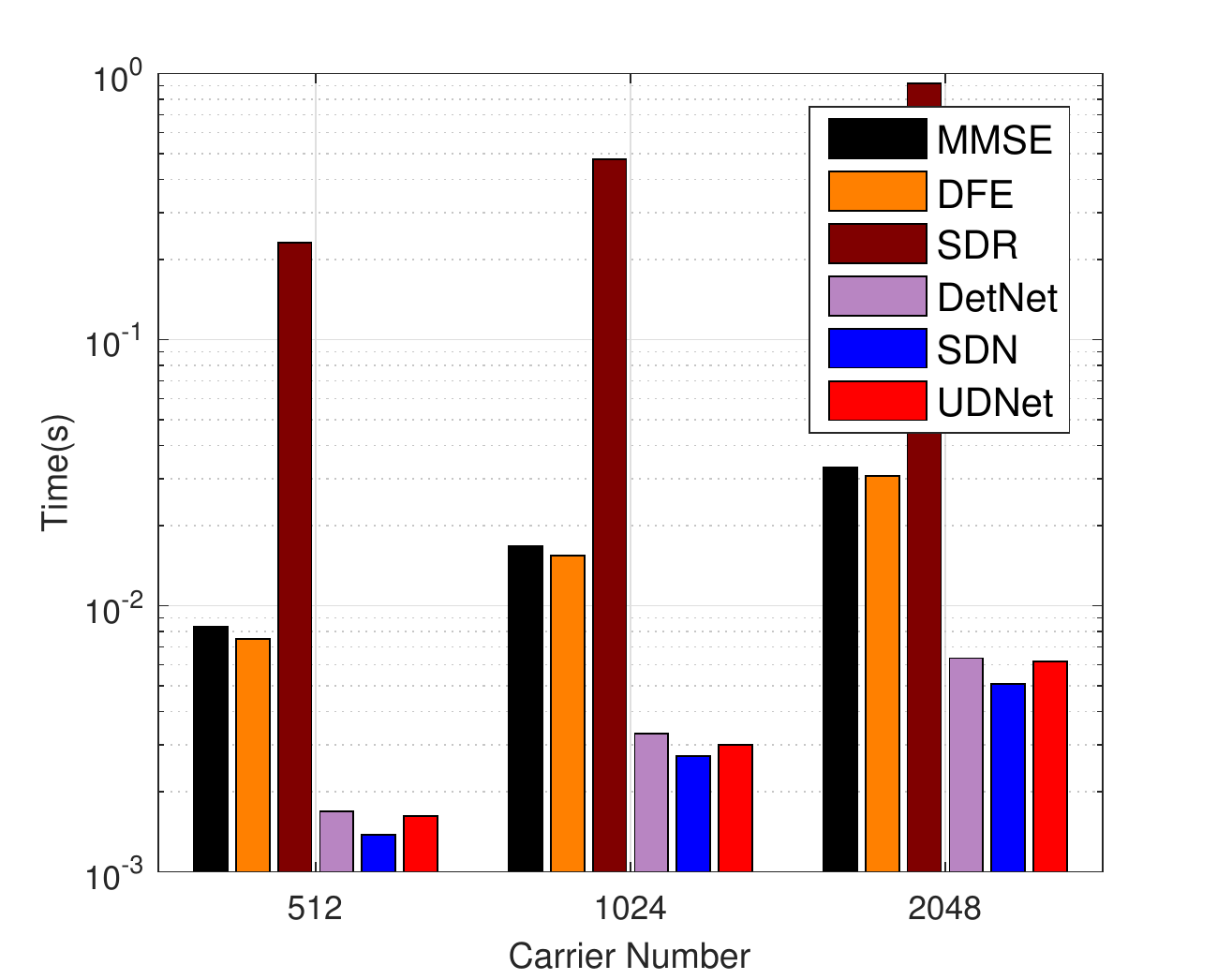}
	\caption{\textcolor{black}{Comparison of the calculation time of different algorithms.}}\label{computingtime}
\end{figure}

\begin{figure}[tb]
	\captionsetup{font={small}}
	\centering 
	\includegraphics[width=9cm]{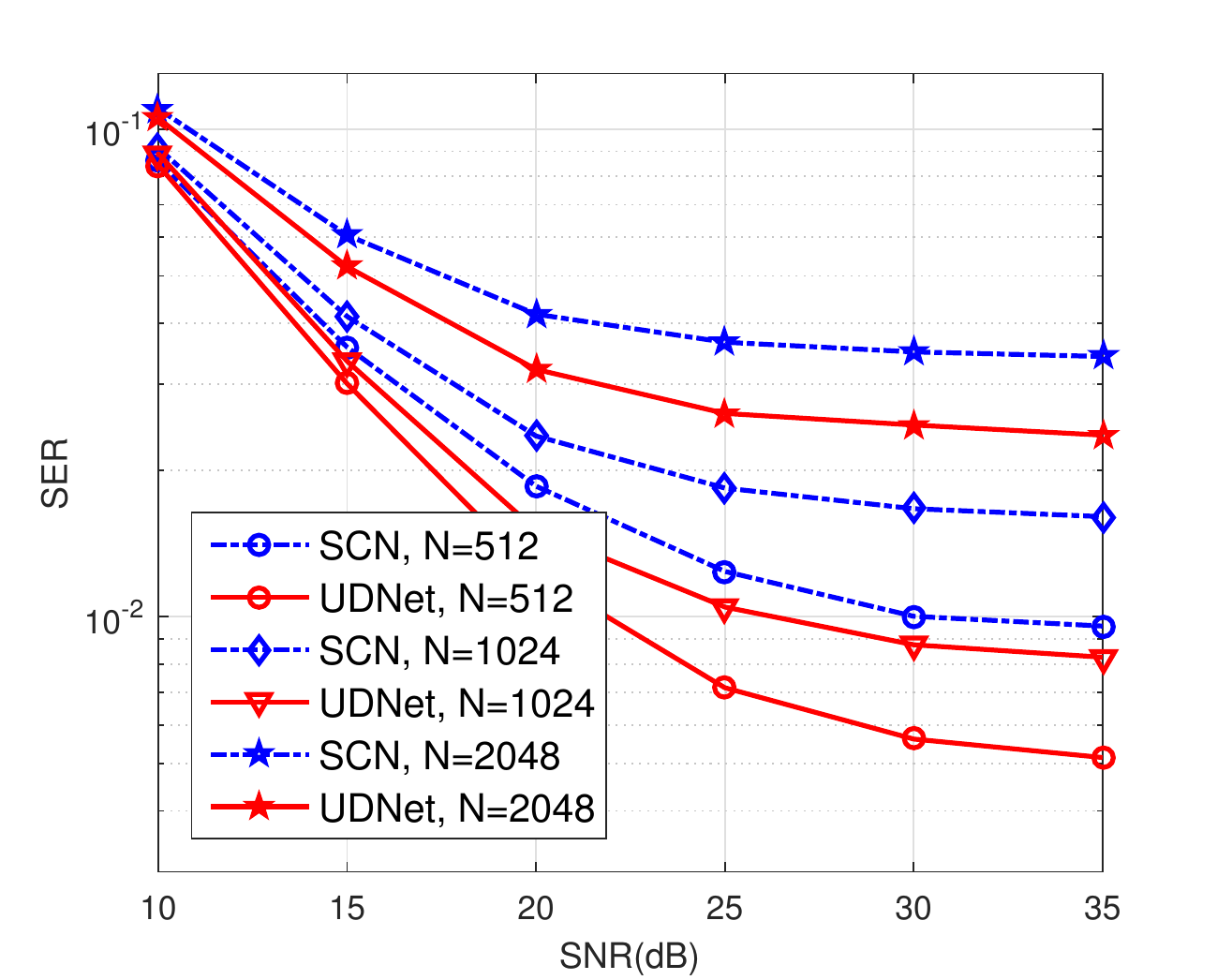}
	\caption{\textcolor{black}{Comparison of SER performance of SCN and UDNet over NOF channel with different subcarrier numbers.}}\label{Fig18_SCNvsUDN}
\end{figure}
\textit{1) Model Size:} \textcolor{black}{
	Table~\ref{tab3} compares the complexities of the equalizers based on deep unfolding parameters, including the amount of floating-point multiplication adds (FLOPs), the number of model parameters, the model size, and the multiplication and addition times of single-layer iterative networks.} The comparison result shows that the model complexity of UDNet is between DetNet and SCN. It is because each layer of UDNet has fewer learning parameters than DetNet, which reduces the amount of calculation. Furthermore, our network is compared with SCNet. It induces more residual connection structures and a classification structure to improve network performance. Therefore, these added functional structures make UDNet more complex than SCN. 

\textit{2) Computation time:} 
\textcolor{black}{
	In general, the computation time is used to compare the complexity of DL algorithms\cite{b15, SuperEst, complexity1}.
	To further explore the complexity of different algorithms, we conduct the following experiments by changing the number of sub-carriers. 
	Fig. \ref{computingtime} shows the average calculation time required by various algorithms in the face of different numbers of sub-carriers. The experimental results tell that the computational complexity of UDNet is less than that of  other algorithms except for SCN. The outcome is consistent with the result in Table~\ref{tab3}. 
	Moreover,  we can find that when the subcarrier number increases, the computation time exponentially rises for classical equalizers. It is because the classical algorithms involve inverse or division operations, which brings a significant computational performance overhead. 
	DL equalizers only deal with multiplication and addition, so that there is no such problem.
	In conclusion,  DL equalizers have a smaller time overhead than classical equalizers when using the same computing equipment.
}

Moreover, Fig. \ref{Fig18_SCNvsUDN} shows that the SER performance of UDNet is better than SCN when the number of sub-carriers is $N=512$, $N=1024$, and $N=2048$. The difference is because, with the carrier number increase in a fixed bandwidth, the data rate rises, and it makes ICI serve and SER decrease.
Therefore, it can be concluded that UDNet achieves a balance of computation time and accuracy.

\section{Concusions}
In this paper, we proposed an unfolding-based UDNet to equalize OFDM signal transmitted over UWA doubly-selective channel. A  classification structure and cross-layer residual connections were employed for the equalizer. We built several UWA experiment channels that are as realistic as possible. Experimental results indicated that the proposed UDNet outperforms classical equalization algorithms and deep unfolding equalizers, i.e., DetNet and SCN, with outstanding computation time. Moreover, our network has a specific generalization in these experiment channels. Meanwhile, UDNet possessed comparatively better robustness against perfect or imperfect CSI. Apart from this, we proved that the proposed network performs better under realistic underwater conditions with offshore background noise and measured channel impulse response.


\vspace{12pt}
\end{document}